\documentclass[aps, prd, twocolumn, superscriptaddress, nofootinbib]{revtex4}
\usepackage{graphicx}
\usepackage{dcolumn}
\usepackage{amssymb}
\usepackage{amsmath,amssymb,amsfonts}
\usepackage{comment}
\usepackage{hyperref}

\newcommand{\be}{\begin{equation}}
\newcommand{\ee}{\end{equation}}
\newcommand{\bea}{\begin{eqnarray}}
\newcommand{\eea}{\end{eqnarray}}
\begin{document}

\title{Observational constraints on dual intermediate inflation}
\author{John D. Barrow}
\email{jdb34@hemes.cam.ac.uk}
\affiliation{DAMTP, Centre for Mathematical Sciences, University of Cambridge,
Wilberforce Rd., Cambridge, CB3 0WA, UK}
\author{Macarena Lagos}
\email{m.lagos13@imperial.ac.uk}
\affiliation{Theoretical Physics, Blackett Laboratory, Imperial College, London, SW7 2BZ,
UK}
\affiliation{Astrophysics, University of Oxford, DWB, Keble Road, Oxford OX1 3RH, UK}
\author{Jo\~{a}o Magueijo}
\email{j.magueijo@imperial.ac.uk}
\affiliation{Theoretical Physics, Blackett Laboratory, Imperial College, London, SW7 2BZ,
UK}
\date{\today }

\begin{abstract}
We explore the observational implications of models of intermediate
inflation driven by modified dispersion relations, specifically those
representing the phenomenon of dimensional reduction in the ultraviolet limit.
These models are distinct from the standard ones because they do not require
violations of the strong energy condition, and this is reflected in their
structure formation properties. We find that they can naturally accommodate
deviations from exact scale-invariance. They also make clear predictions for
the running of the spectral index and tensor modes, rendering the models
straightforwardly falsifiable. We discuss the observational prospects for
these models and the implications these may have for quantum gravity
scenarios.
\end{abstract}

\maketitle

%=====================================================================
%=====================================================================
%=====================================================================

\section{Introduction}

The phenomenon of dimensional reduction in the ultraviolet (UV) energy
limit has attracted a growing interest from the quantum gravity community 
\cite%
{Lollprl,Litim,Reuter,Hl,HLspec,Alesci:2011cg,Benedetti:2008gu,Modesto1,Caravelli,Magliaro,Modesto2,Calcagni1,Calcagni2}%
. In some of these scenarios it appears that modified dispersion relations
(MDRs) encode the salient features of the phenomenon \cite{visser}.
Interestingly, these modified dispersion relations may be implemented with 
\cite{Hl,visser}, or without, violating the principle of relativity of
inertial frames \cite{dsrrsd}. The phenomenological implications of the deep
UV limit of these MDRs can be studied by building cosmological models based
upon them, and evaluating their cosmic structure formation properties \cite%
{dimred} so that their predictions can be compared with observations. A
minimal assumption in these calculations is that Einstein gravity (GR) is
valid in the frame where the MDRs are postulated. This is called the
Einstein frame. However, by disformally transforming to a frame which
trivialises the MDRs, we find an equivalent dual description which displays
modified gravity (more specifically 'rainbow gravity' \cite{rainbowDSR}) and
unmodified dispersion relations \cite{rainbowred}. This is called the
'rainbow frame'.

In the Einstein frame, MDRs drive cosmic structure formation \cite%
{csdot,bim,Mag} without the need for inflation, but in the dual frame the
cosmological expansion is perceived as accelerating. Yet, such superluminal
expansion is never conventional inflation and, in particular, it is not
derived from the behaviour of scalar fields or other sources of a violation
of the strong energy condition. The accelerated expansion is driven purely
by modified gravity \cite{Garattini,rainbowred} and, unsurprisingly, the
conditions for a scale-invariant spectrum of density perturbations to arise
are very different to those required in standard inflationary models \cite%
{Starob}.

It is known that power-law MDRs lead to dual power-law and de Sitter inflation,
so one may wonder what type of MDRs dualize into intermediate inflation \cite%
{intinfl0,intinfl1,intinfl2,intinfl3,intinfl4} -- another example of the
inflationary scenario. In \cite{inter-rainb} it was shown that to achieve
this style of inflation it was enough to modulate the power-law appearing in
the MDRs with a logarithmic factor, typical of those arising from
renormalization arguments. The purpose of this paper is to study the full
observational constraints upon this model, in the light of recent
observational results \cite{Planck}. We will also study how these
constraints may feed back into quantum gravity theories, and bring a much
larger data set to bear upon them.

We start by presenting the model and improving the presentation in \cite%
{inter-rainb}. We shall write the MDRs, proposed in \cite{inter-rainb}, in the
alternative form: 
\begin{equation}
E^{2}=p^{2}\left[ 1+\left( \lambda p\right) ^{2\gamma }\left( \frac{\ln
(p_{0}/p)}{\ln (\lambda p_{0})}\right) ^{2\beta }\right],  \label{Mdr}
\end{equation}
(valid for massless particles, but easy to adapt for massive particles) with 
$\beta$, $\gamma$, $\lambda$ and $p_{0}$ non-negative constants. The speed of light $c$ is
obtained from the MDR via $c=dE/dp\approx E/p$. This reparameterisation has
two advantages. First, it brings to the fore the fact that there is a
maximum momentum $p_{\text{M}}$ in these models, as already pointed out in 
\cite{inter-rainb}. The maximum momentum is more precisely defined as the
point where $c=dE/dp=0$ and this is only of the order of $p_{0}$ (for the
cases we are interested in; with $\gamma $ and $\beta $ of the order of
unity); specifically, $p_{\text{M}}=p_{0}\exp \left( \frac{-\beta }{1+\gamma 
}\right) $.  Note that $p_0$ will always have to be a very large number, so that the cut off 
happens at energies higher than those needed to solve the horizon problem. This is 
a general feature of these models, unrelated to the issue of the fine tuning necessary for
obtaining departures from scale-invariance, which is the subject of this paper. 

The second advantage  of parameterisation (\ref{Mdr}) concerns the logarithmic factor introduced
in the denominator. With this factor, the transition between the infrared (IR) regime and the UV regime is always at 
$p\sim \lambda ^{-1}$ (regardless of the remaining parameters). The IR regime is described then by $\lambda p\ll 1$, where
the MDR turns out to be trivial as $E^{2}\approx p^{2}$, while in the UV
regime, $\lambda p\gg 1$, there is a strong modification: $E^{2}\approx
p^{2}\left( \lambda p\right) ^{2\gamma }\left( \ln (p_{0}/p)/\ln (\lambda
p_{0})\right) ^{2\beta }$. This will clarify some of the points we will make
regarding the spectrum normalization, and the implications for quantum
gravity theories.

We highlight the specific case $\gamma =2$, known to be related to conformal
invariance of the gravitational coupling \cite{rainbowred} as well as to
strict scale-invariance for the fluctuations. The logarithmic factor can
then be seen as a soft breaker of conformal invariance. The appearance of
such soft breaking is to be expected in any theory as was pointed out at the
end of \cite{measure}. Whether the resulting departure from strict
scale-invariance can \textit{naturally} accommodate the observations is one
of the questions raised by this paper. We will argue that it can.

The plan of the rest of this paper is as follows. In section \ref{SectionDim}
we study the model in the context of quantum gravity, and the phenomenon of
dimensional reduction in the UV regime. Specifically, we present the running
spectral and Hausdorff dimensions for the most relevant case, with $\gamma
=2 $ and $\beta =1$. Then, in Section \ref{SectionPert} we find the power
spectrum for scalar and tensor perturbations in general, and focus later on
the specific case $\gamma =2$. We find that for $\beta \neq 0$, this theory
naturally predicts deviations from exact scale-invariance at the rough level
required by observations. In Section \ref{SectionObs}, we compare the
predictions with the observations in more detail, and derive constraints
upon the parameters of the theory. In a concluding section we examine what
the wider implications might be for quantum gravity theories.

Throughout this paper we will use Planck units (i.e.~$c=\hbar=G=1$).

\section{Dimensional reduction associated with the MDRs}

\label{SectionDim}

In this section we work out the dimensional reduction profile linking the IR
regime and the UV regime associated with (\ref{Mdr}). We perform this using both the
spectral dimension measure and the Hausdorff dimension of momentum space in
a dual picture where the MDRs are trivialised, with the measure absorbing
the non-trivial effects. A number of concerns have been raised regarding the
use of the spectral dimension and its probabilistic interpretation \cite{astrid}.
These are beyond the scope of this paper, and in any case we find equivalent
results whatever dimensionality measure we use, as we will now show.

\subsection{The spectral dimension}

The spectral dimension $d_{s}(s)$ has been proposed as a possible quantity
characterising the geometry of some quantum gravity theories (\cite%
{Ambjorn:2005db}). We can think of the spectral
dimension as the effective dimension probed by a fictitious random walk
process. Its average return probability at a scale $s$ is given by: 
\begin{equation}
P(s)=\int \frac{d^{3}pdE}{(2\pi )^{4}}e^{-s\Omega _{\lambda }(p,E)},
\label{Ps}
\end{equation}%
where $\Omega _{\lambda }(p,E)=E^{2}+f_{\lambda }(p)$, and the defining
function $f_{\lambda }(p)$ is supplied by the specific MDR. In our case, we
have 
\begin{equation}
f_{\lambda }(p)=p^{2}\left[ 1+(\lambda p)^{2\gamma }(\ln (p_{0}/p)/\ln
(\lambda p_{0}))^{2\beta }\right] .
\end{equation}
We define the spectral dimension as: 
\begin{equation}
d_{s}(s)=-2\frac{d\ln P(s)}{d\ln s}.
\end{equation}%
The parameter $s$ can be interpreted as the scale at which we are probing
the process. For $s\rightarrow 0$ we probe the ultraviolet limit, while for $%
s\rightarrow \infty $ we probe the infrared. In general, the integral in (%
\ref{Ps}) should consider all the possible values of the energy $E$ and
momentum $p$. This means that, for our MDR, the integral of momentum must be
done up to its maximum, $p_{\text{M}}$.

As has been mentioned in \cite{dimred,inter-rainb}, the case of $\gamma =2$
is of special interest. This case leads to exact scale-invariance for $\beta
=0$ (which describes de-Sitter spacetime in the rainbow frame), and leads to
a UV spectral dimension of about $2$, which is favoured in many
quantum-gravity studies (\cite{Ambjorn:1997jf, Horava:2009if, Ambjorn:2005db, Lauscher:2005qz, Kommu:2011wd, Giasemidis:2012rf, Alesci:2011cg, Modesto:2008jz, Calcagni:2010pa}). For this reason, we will study that particular case. Fig.~\ref%
{SpecDimfig1} shows the result calculated for the spectral dimension $%
d_{s}(s)$ in the case of $\lambda =1$, $p_{0}=10^{50}$, $\gamma =2$ and $%
\beta =1$, in the region where the transition from UV to IR occurs.

\begin{figure}[h]
\begin{center}
\scalebox{0.53}{\includegraphics{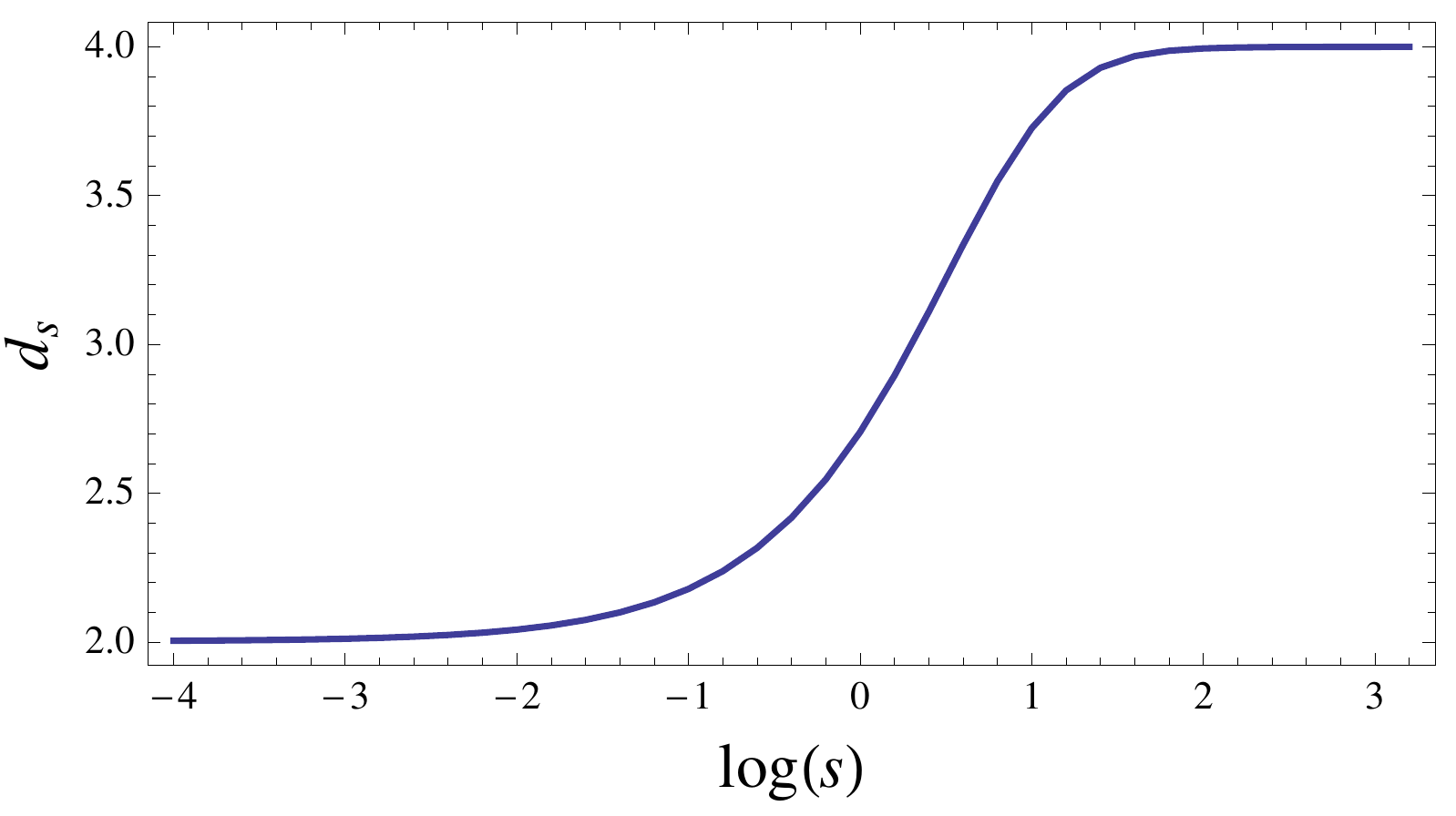}}
\end{center}
\caption{The transition from UV to IR of the spectral dimension, $d_{s}(s)$
for $\protect\lambda =1$, $p_{0}=10^{50}$, $\protect\gamma =2$ and $\protect%
\beta =1$.}
\label{SpecDimfig1}
\end{figure}

From Fig.\ref{SpecDimfig1} we can see that in the IR regime, $%
\lim_{s\rightarrow \infty }d_{s}=4$, which is expected since eq.~(\ref{Mdr})
approaches the trivial dispersion relation $E^{2}=p^{2}$ in this regime, and 
$d_{s}$ then coincides with the topological dimension. In addition, we can
see that in the UV regime, $d_{s}$ is near to 2, which looks similar to what
was found in \cite{dimred} for $\gamma =2$ and $\beta =0$. However, Fig.~\ref%
{SpecDimfig2} shows that these cases are actually different. In \cite{dimred}%
, it was found that $\lim_{s\rightarrow 0}d_{S}=2$, but now we observe that $%
d_{s}$ never settles to 2, but it keeps on increasing as $s\rightarrow 0$. 
% As a matter of fact, we are not even able to look at the case of $s\rightarrow 0$, because the maximum momentum $p_\text{M}$ translates into a minimum $s$. 

\begin{figure}[h]
\begin{center}
\scalebox{0.53}{\includegraphics{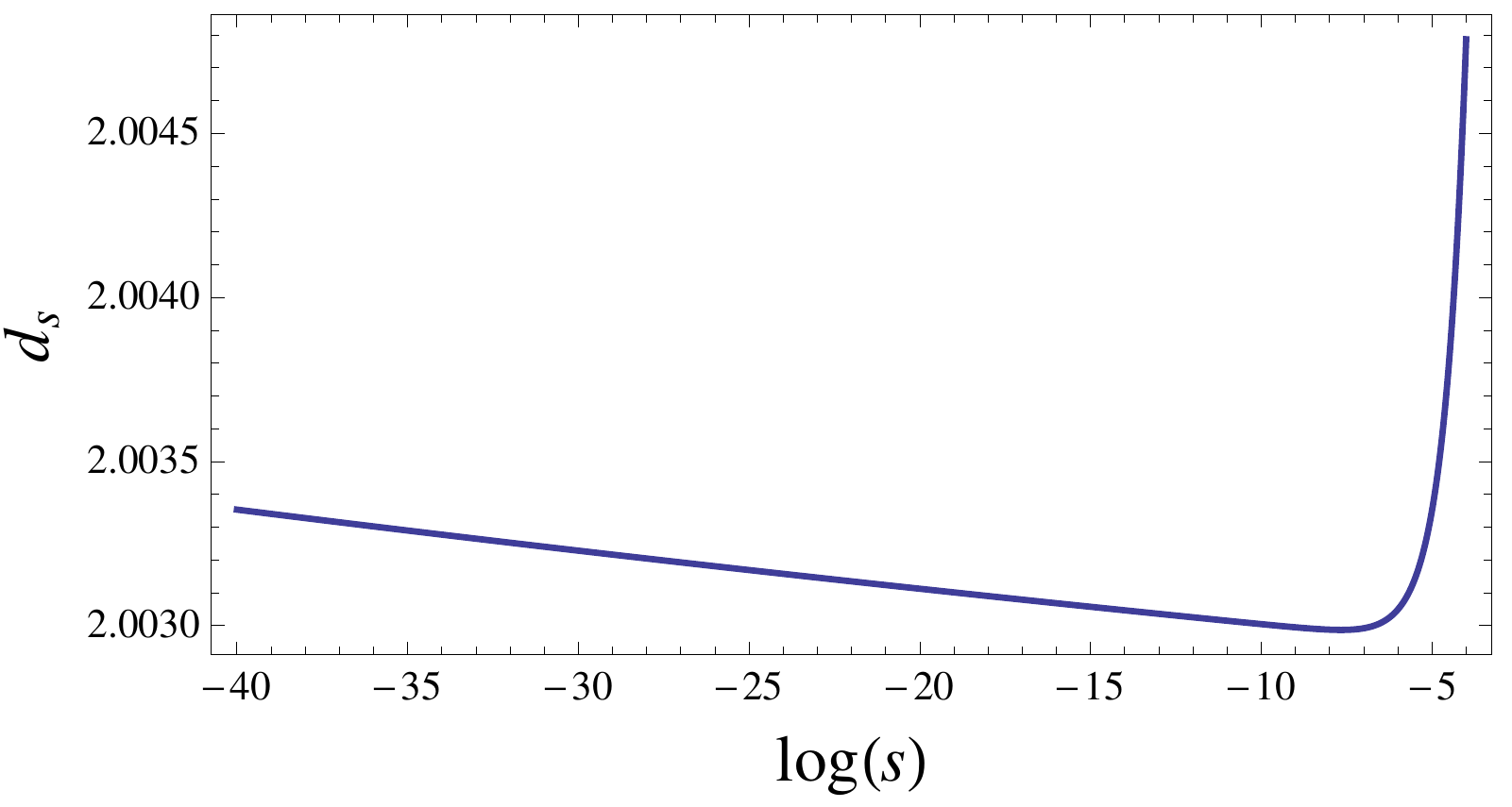}}
\end{center}
\caption{The UV region of the spectral dimension variation with $s$ for $%
\protect\lambda =1$, $p_{0}=10^{50}$, $\protect\gamma =2$ and $\protect\beta %
=1$. Note that $d_{s}$ never settles at 2, but keeps on increasing as $%
s\rightarrow 0$.}
\label{SpecDimfig2}
\end{figure}

\subsection{The dual Hausdorff dimension}

An alternative characterisation of the phenomenon of dimensional reduction
was proposed in \cite{measure}. By redefining the units of momentum it is
possible to trivialise the dispersion relations. This shifts the non-trivial
effects elsewhere, for example to the interactions, or to the measure of
integration in momentum space. Interestingly, the latter shows a Hausdorff
dimension $d_{H}$ which in the UV limit coincides with the UV spectral
dimension of the theory. It was shown in \cite{dsrrsd} that a similar
phenomenon may be found in theories which do not introduce preferred frames,
but the integration measure in energy-momentum space becomes non-factorable.

\begin{figure}[h]
\begin{center}
\scalebox{0.53}{\includegraphics{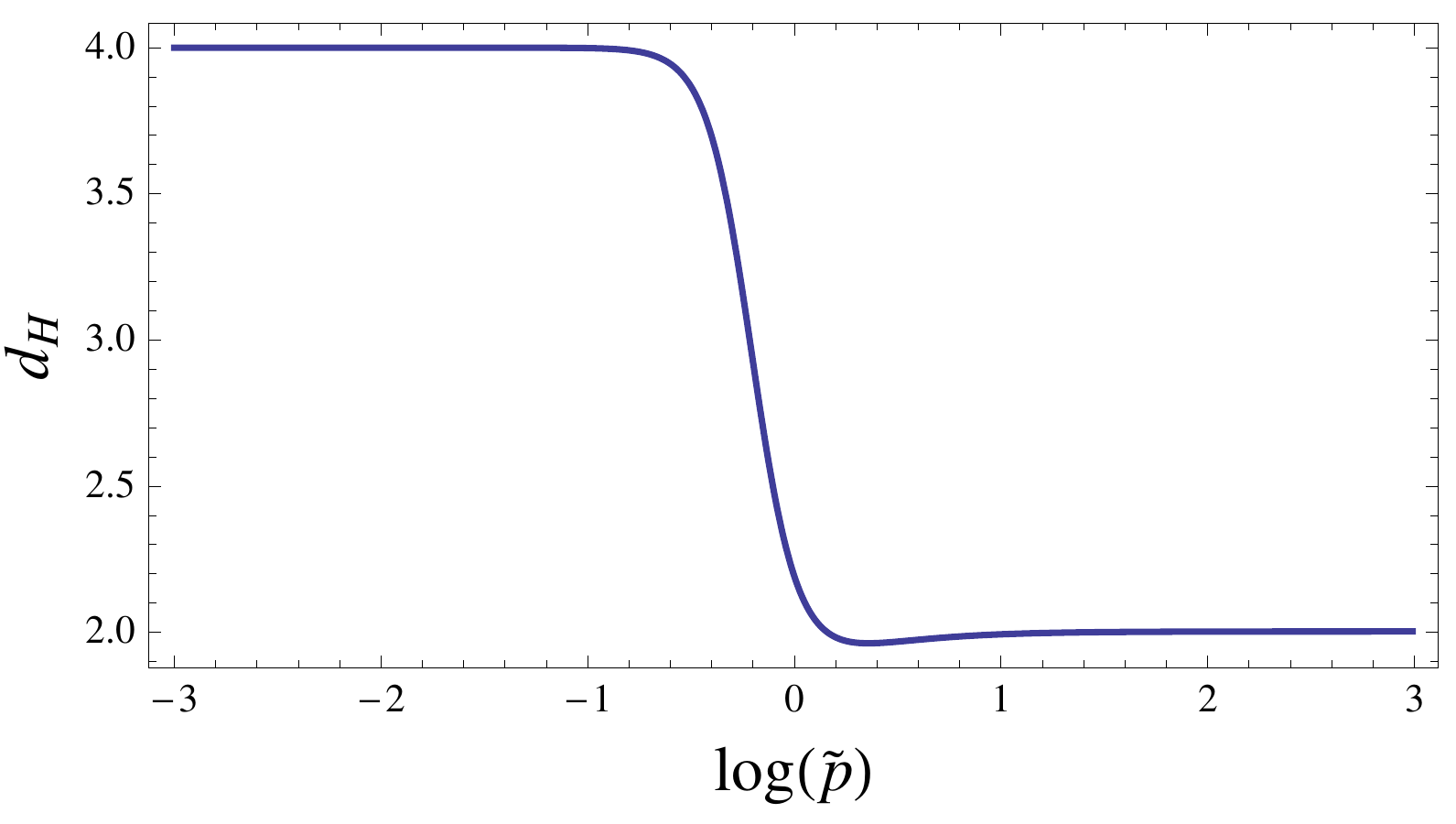}}
\end{center}
\caption{The transition from IR to UV of the Hausdorff dimension, $d_{H}(%
\tilde{p})$, of momentum space for $\protect\lambda =1$, $p_{0}=10^{50}$, $%
\protect\gamma =2$ and $\protect\beta =1$.}
\label{hausfig1}
\end{figure}

In our case, trivialising the MDR can be done by defining a new (spatial)
momentum variable: 
\begin{equation}
\tilde p = p \left[1+(\lambda p)^{2\gamma}\left(\frac{\ln (p_0/p)}{ \ln
(\lambda p_0)}\right)^{2\beta}\right]^{1/2}\, .
\end{equation}
We can then evaluate the momentum measure $d\mu({\tilde p})=\mu({\tilde p}%
)d\tilde p=p^2dp$, and evaluate its running Hausdorff dimension from: 
\begin{equation}
d_{H}({\tilde p})=2+ \frac{d\ln\mu}{d\ln \tilde p}.
\end{equation}
Fig.~\ref{hausfig1} shows the results for the Hausdorff dimension with $%
\lambda =10^{4}$, $p_{0}=10^{50}$, $\gamma =2$ and $\beta =1$, during the
transition from IR to UV. We observe a similar transition behaviour to that
found for the spectral dimension.

\begin{figure}[h]
\begin{center}
\scalebox{0.53}{\includegraphics{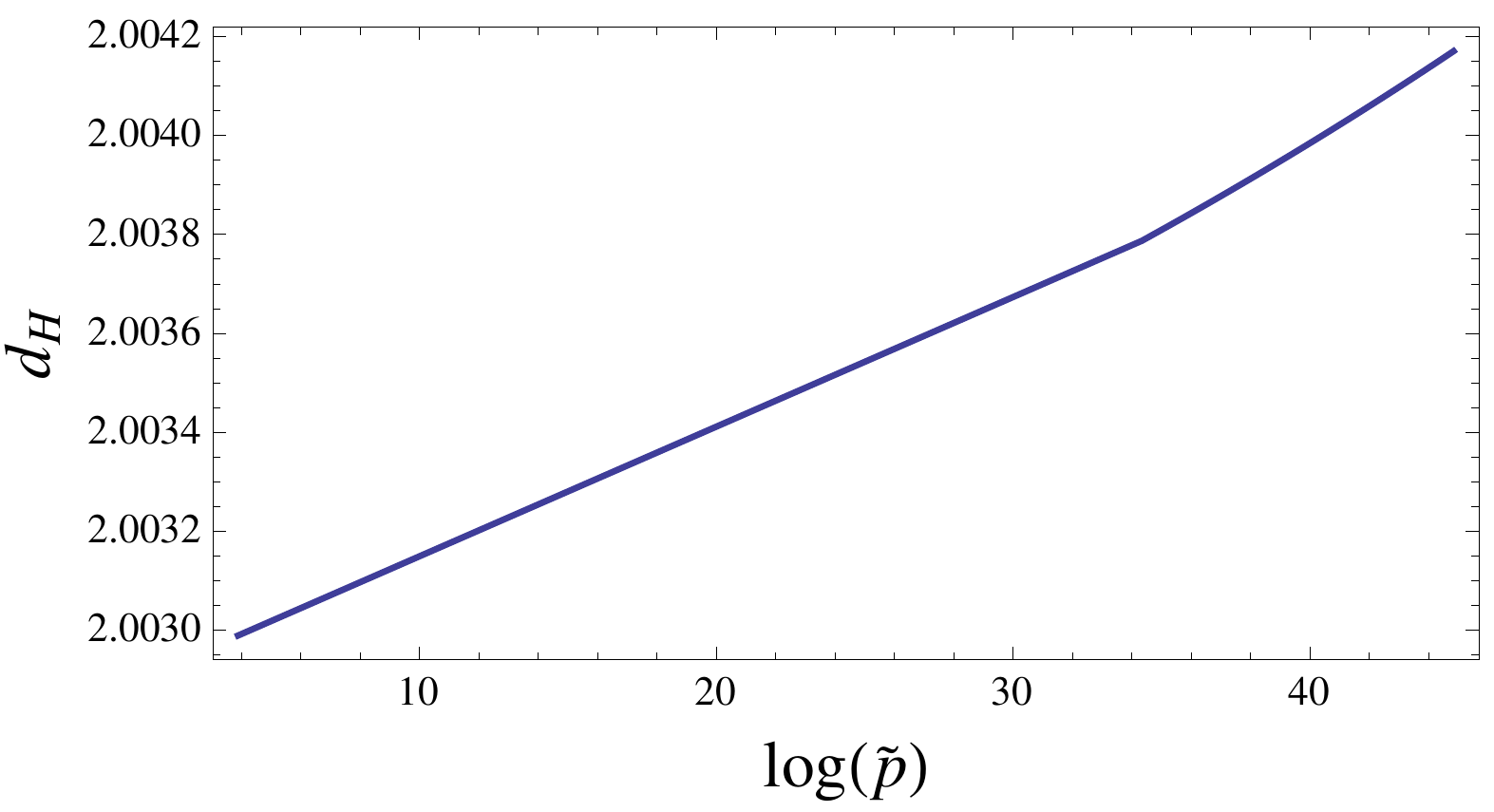}} \scalebox{0.53}{%
\includegraphics{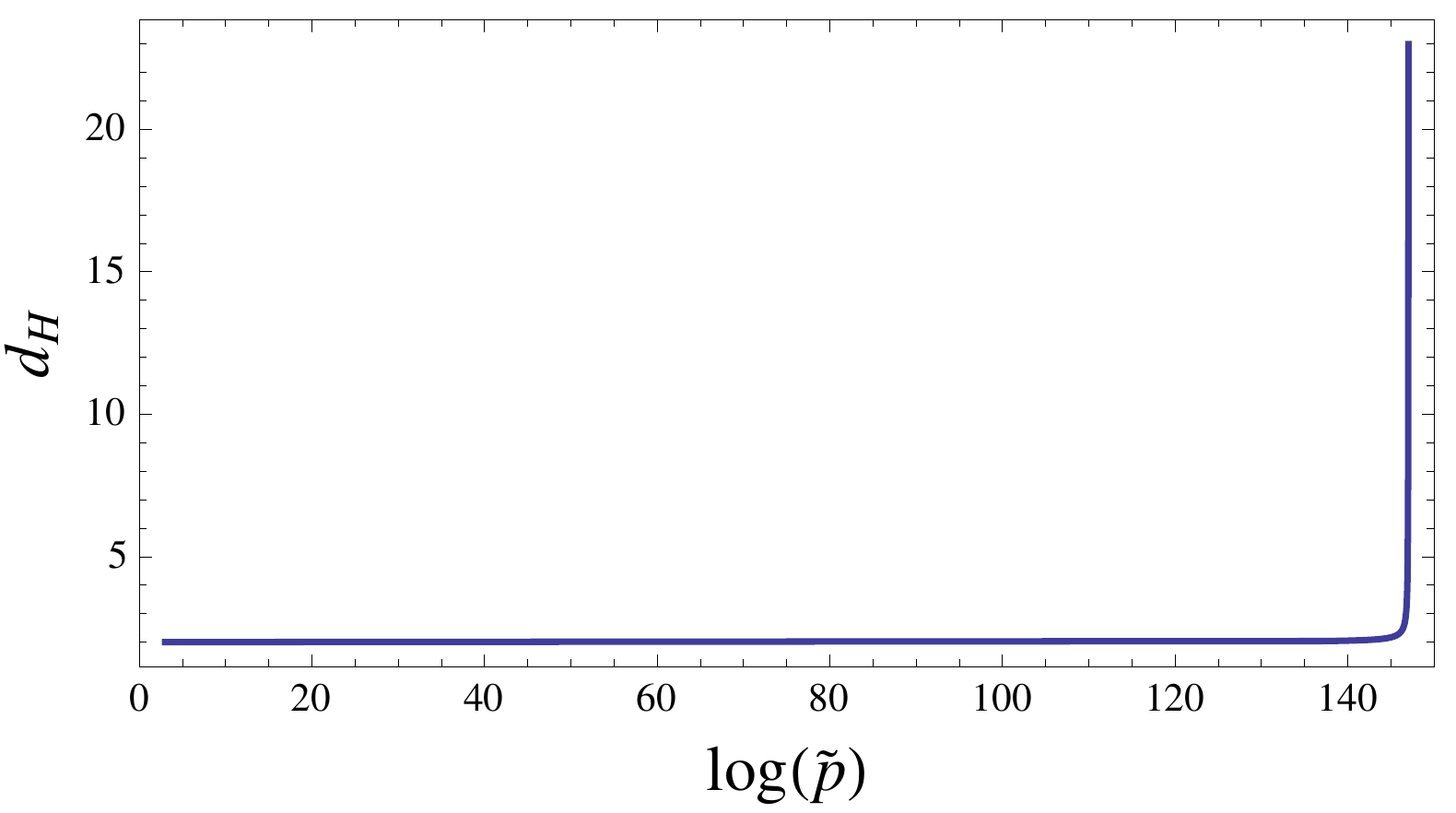}}
\end{center}
\caption{The UV region of the Hausdorff dimension of momentum space for $%
\protect\lambda =1$, $p_{0}=10^{50}$, $\protect\gamma =2$ and $\protect\beta %
=1$. The top plot shows some UV region, where it is seen that $d_{H}$ never
settles at 2, but it rather keeps on increasing. The bottom plot, with
greatly expanded scales, shows the long UV regime, where a divergence is
found at a maximum momentum $\tilde{p}_{M}$.}
\label{hausfig2}
\end{figure}

In addition, Fig.~\ref{hausfig2} shows the UV region. The top plot shows the
same behaviour as in the spectral dimension, where $d_{H}$ never settles to
2. However, in this case there is a maximum momentum $\tilde{p}_{\text{M}}$
(corresponding to $p_{\text{M}}$), where $d_{H}\rightarrow \infty $, as it
can be seen in the bottom plot of this figure with greatly expanded scales.

Notice that these results for the spectral and Hausdorff dimensions are
general for $\gamma =2$. The specific value of $\beta $ does not affect the
behaviour of $d_{s}$ nor $d_{H}$ significantly (except when $\beta =0$)
because $\beta $ can only have an effect in the UV limit but in this case
the polynomial term $\lambda p$ with exponent $\gamma $ dominates over the
logarithmic term.

\section{Detailed calculation of the cosmological scalar and tensor
fluctuations}

\label{SectionPert}

MDRs models are able to produce viable primordial perturbations in the early
Universe, providing an alternative to the simple models of inflation (\cite%
{Dodelson:2003}). This has been already shown in \cite%
{dimred,Mag,inter-rainb,Lu:2009he}.

In this section we calculate the scalar and tensor power spectra predicted
with the MDR given by eq.~(\ref{Mdr}). We will find their amplitudes along
with the spectral indices, and due to the special interest in the case with $%
\gamma =2$, we will specialise the calculations to that case at the end.

\subsection{Primordial scalar fluctuations}

Let us assume that the underlying theory is GR (see \cite{Mag,dimred} for
the logic behind this choice) and consider the equation of
motion for first-order scalar cosmological perturbations in a spatially flat
Friedmann-Robertson-Walker (FRW) universe filled with a perfect fluid: 
\begin{equation}  \label{EqPert}
v^{^{\prime \prime }}+\left(c^2k^2-\frac{a^{^{\prime \prime }}}{a}\right)v=0,
\end{equation}
such that $v=-\zeta /a$, and $c=E/p$ (up to a factor of order 1). We are
ultimately interested in calculating the field $\zeta$, known as the
curvature perturbation. The general form of this equation of motion can be
found in \cite{Mukhanov:1990me}, but we have already made some
approximations given our MDR \cite{Mag,dimred}.

For an equation of state of matter $P=w\rho $ (with $P$ being the pressure
and $\rho $ the energy density), for a spatially-flat FRW universe we
have that: 
\begin{equation}
a\propto \eta ^{\frac{1}{\epsilon -1}};\quad \epsilon =\frac{3}{2}(1+w),
\label{aeta}
\end{equation}%
where $\eta $ is the conformal time satisfying $\eta >0$, and $w$ is assumed to be
constant. In this model, the horizon problem can be solved in an expanding Universe that does not necessarily inflate, i.e.~with $w>-1/3$.
This is because if we take in consideration eq.~(\ref{Mdr}), in the past, i.e.~in the UV regime, $c$ evolves in time as:
\be
 c\propto \eta^{-\alpha}[C_1+C_2\ln(\eta)]^\beta ; \quad \alpha=\frac{\gamma}{\epsilon-1},
\ee
where $C_1$ and $C_2$ are some constants.
Since we also have $a''/a\propto 1/\eta ^{2}$, and we will be considering only cases where $-1/3<w<(2\gamma-1)/3$ (and therefore $\alpha>1$), then $c^2k^2\eta^2\ll 1$ in this regime. This means that the first term in the parenthesis of (\ref{EqPert}) dominates over the second term. Consequently, we have that in the past perturbations oscillate on sub-Hubble scales ($ck\eta \gg 1$). As time grows the second term dominates, i.e.~perturbations freeze-in on super-Hubble scales ($ck\eta \ll 1$). Due to the fact that all perturbations were inside the Hubble radius during the UV regime, the horizon problem is solved.

Now, we would like to solve eq.~(\ref{EqPert}), but instead of solving it
exactly, we do so in two regimes: for sub-Hubble and super-Hubble scales,
and then match both solutions at the horizon-crossing time. For sub-Hubble
scales, the correct normalised solution is given by the WKB
approximation: 
\begin{equation}
v\approx \frac{e^{ik\int c\,d\eta }}{\sqrt{ck}}.
\end{equation}%
For super-Hubble scales, the solution is simply given by $%
v=F(k)a$, where $F(k)$ is some undetermined function. Next, we impose a
continuity condition on $v$ at the horizon-crossing time $\eta_*$ defined by $c_{\ast }k\eta
_{\ast }=1$, in order to find $F(k)$, which actually corresponds to $\zeta $
on super-Hubble scales. For simplicity, in the following calculations we
will use the parametrisation of the MDR given in \cite{inter-rainb}: 
\begin{equation}
E^{2}=p^{2}\left( 1+(\tilde{\lambda}p)^{2\gamma }(D-\ln (\tilde{\lambda}%
p))^{2\beta }\right) ,
\end{equation}%
where the relation between the parameters $(D,\tilde{\lambda})$ and $%
(p_{0},\lambda )$ of (\ref{Mdr}) is: 
\begin{equation}
D=\ln \left( \frac{\lambda p_{0}}{\ln (\lambda p_{0})^{\beta /2}}\right)
;\quad \tilde{\lambda}=\frac{\lambda }{\ln (\lambda p_{0})^{\beta /2}}.
\label{RelParam}
\end{equation}%
By assuming that the crossing-time occurs in the UV regime, which as we will
see later can be guaranteed by choosing a suitable value of $\lambda $, we
can approximate $c_{\ast }\approx (\lambda p_{\ast })^{\gamma }(D-\ln
(\lambda p_{\ast }))^{\beta }$ and find that the power spectrum of $\zeta $
on super-Hubble scales is: 
\begin{equation}
\mathcal{P}_{\zeta }(k)=\frac{k^{3}}{2\pi ^{2}}|\zeta |^{2}\sim \frac{\left( 
\tilde{\lambda}^{\gamma }D^{\beta }\right) ^{\frac{\gamma -2}{\gamma
+1-\epsilon }}k^{\frac{\epsilon \left( \gamma -2\right) }{\gamma +1-\epsilon 
}}}{2\pi ^{2}\tilde{\lambda}^{\gamma }\left[ E+\left( \frac{\gamma +1}{%
1+\gamma -\epsilon }-1\right) \ln (\tilde{\lambda}k)\right] ^{\beta }},
\label{PowSpec}
\end{equation}%
such that $E\equiv D+\ln \left[ \left( D^{\beta }/\tilde{\lambda}\right) ^{%
\frac{1}{1+\gamma -\epsilon }}\right] $.

In order to see more clearly the dependence of the power spectrum on $k$, we
approximate the power spectrum on some scale $k_{0}$ by 
\begin{equation}
\mathcal{P}_{\zeta }(k)\sim A_{\zeta }^{2}\left( \frac{k}{k_{0}}\right) ^{n_{%
\text{s}}(k_{0})-1},  \label{PS1}
\end{equation}%
where $A_{\zeta }^{2}=\mathcal{P}_{\zeta }(k_{0})$ is the amplitude of the
power spectrum at $k_{0}$, and $n_{\text{s}}$ is the spectral index given
by: 
\begin{align}
& n_{\text{s}}(k)-1=\frac{d\ln \mathcal{P}_{\zeta }(k)}{d\ln k}
\nonumber\\
& =\frac{\epsilon \left( \gamma -2\right) }{1+\gamma -\epsilon }-\frac{\beta
\epsilon }{\left( 1+\gamma -\epsilon \right) \left( E+\left( \frac{\gamma +1%
}{1+\gamma -\epsilon }-1\right) \ln (\tilde{\lambda}k)\right) }\label{SpecIn} 
\end{align}%
Here, we can see two terms: the first comes from the polynomial part of the
dispersion relation; it was already found in \cite{Mag}, and corresponds to
the particular case of $\beta =0$. The second term comes from the
logarithmic part of (\ref{Mdr}) and brings a dependence on the scale $k$
into the spectral index, so~the power spectrum cannot be written exactly as
a purely polynomial function of $k$.

Now, we consider the specific case of $\gamma =2$. The power spectrum (\ref%
{PowSpec}) and the spectral index (\ref{SpecIn}) reduce to: 
\begin{equation}
\mathcal{P}_{\zeta }\sim \frac{1}{2\pi ^{2}\tilde{\lambda}^{2}(E+2\ln (%
\tilde{\lambda}k))^{\beta }};\quad E=D+\ln (D^{\beta }/\tilde{\lambda}),
\label{PSt}
\end{equation}%
and 
\begin{equation}
n_{\text{s}}(k)-1=-\frac{2\beta }{E+2\ln (\tilde{\lambda}k)}.
\label{SpecIn2}
\end{equation}%
Here, we have also set $\epsilon =2$ ($w=1/3$, i.e.~radiation) since, as it
was shown in \cite{inter-rainb}, in order to have a transformation from the
Einstein frame with the MDR (\ref{Mdr}) to a rainbow frame with intermediate
inflation and a trivial dispersion relation, we need that $\gamma =\epsilon $%
. In this rainbow frame, the scale factor evolves as: 
\begin{equation}
a(t)\propto e^{At^{n}},
\end{equation}%
where $n$ is related to the exponents in the MDR by: 
\begin{equation}
\beta =\frac{1}{n}-1.
\end{equation}

From eq.~(\ref{SpecIn2}) we can see that we will always have a deviation
from scale-invariance as long as $\beta \neq 0$, i.e.~in all the possible
models of intermediate inflation. Therefore, the logarithms in eq.~(\ref{Mdr}%
) break exact scale-invariance in the cosmological density fluctuation
spectrum for $\gamma =2$.

\subsection{Primordial tensor fluctuations}

The equation of motion for the tensor modes, described by the field $h$, is
such that if $\tilde{h}=ah$ then $\tilde{h}$ satisfies the same equation as $%
v$, eq.~(\ref{EqPert}). Therefore, we should find the same result for $h$ as
for $\zeta $. However, as was pointed out in \cite{dimred}, the MDR for
gravity and for matter do not need to be the same, and consequently the
expression for $c$ in the equation of motion could now be different. One
simple modification of the MDR in eq.~(\ref{Mdr}) could be: 
\begin{equation}
E^{2}=p^{2}\left[ 1+b^{2}\left( \lambda p\right) ^{2\gamma }\left( \frac{\ln
(p_{0}/p)}{\ln (\lambda p_{0})}\right) ^{2\beta }\right] ,  \label{Mdr2}
\end{equation}%
where $b$ is some dimensionless factor, whose effect is to create a constant
difference in the UV regime between the speeds of gravity and light: 
\begin{equation}
c_{\text{g}}=bc.
\end{equation}%
Note that this case is equivalent to having a different $\lambda $ factor
for tensor perturbations. When this is the case, the power spectrum for
tensor perturbations, $\mathcal{P}_{\text{T}}(k)$, at the horizon-crossing
time is: 
\begin{equation}
\mathcal{P}_{\text{T}}(k)=\left( \frac{2}{b\pi ^{2}}\right) \frac{1}{\tilde{%
\lambda}^{2}(E+\ln (b)+2\ln (\tilde{\lambda}k))^{\beta }},
\end{equation}%
with a tensor spectral index given by: 
\begin{equation}
n_{\text{T}}(k)=-\frac{2\beta }{E+2\ln (\tilde{\lambda}k)+\ln (b)};\;%
\mathcal{P}_{\text{T}}\sim A_{\text{T}}^{2}\left( \frac{k}{k_{0}}\right)
^{n_{T}(k_{0})}.
\end{equation}%
Thus, the tensor-to-scalar ratio will be: 
\begin{equation}
r=\frac{\mathcal{P}_{\text{T}}}{\mathcal{P}_{\zeta }}=\frac{4}{b}\left( 1+%
\frac{\ln (b)}{E+2\ln (\tilde{\lambda}k)}\right) ^{-\beta }.
\end{equation}

The results found in this section for primordial perturbations are valid
while they are frozen-in, i.e.~until they re-enter the horizon again. Since
for a given comoving momentum, $k$, the end of the varying-$c$ period ($%
\lambda p\sim 1$) happens when $ck\eta \sim k^{2}\lambda $ (ignoring the
matter epoch for simplicity, and defining $a=1$ today), then for all
cosmologically relevant scales we have a set of suitable $\lambda $ values
such that $k^{2}\lambda \ll 1$, i.e.~the speed of light becomes constant
while the perturbations are on super-Hubble scales. This means that
perturbations will re-enter the horizon when $c=1$, which lets us compare
our results to the observational ones. In addition, this also means that
perturbations crossed the horizon for the first time during the UV regime,
as we assumed in the previous calculations.

\section{Observational constraints when $\boldsymbol{\protect\gamma =2}$}

\label{SectionObs}

%Before constraining our results with experiments we must first notice that since there is a maximum momentum $p_\text{M}$ for which this theory is valid, we need a consistency condition in order to be able to explain observations. This condition is that all cosmologically relevant scale today must have been gone through the process of being a sub-Hubble scale and then crossing the horizon. For the case of $\gamma=\epsilon=2$, $\lambda\sim 10^4$, this impose a minimum of roughly $ p_0\sim 10^{110}$. 

\begin{figure}[h]
\begin{center}
\scalebox{0.43}{\includegraphics{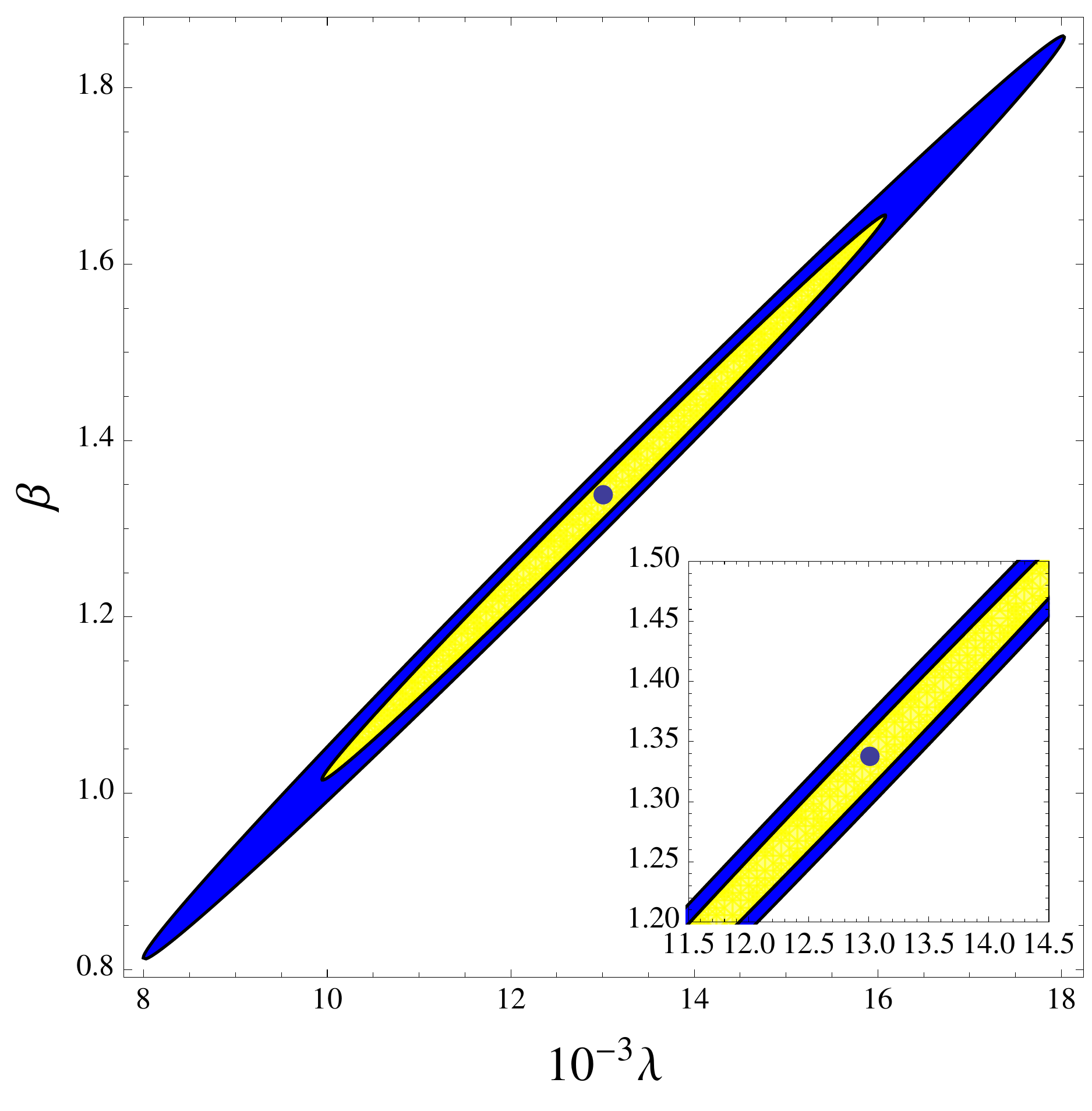}}
\end{center}
\caption{The two-parameter $(\protect\beta ,\protect\lambda )$ joint error
bounds for $p_{0}=10^{135}$ and $\protect\gamma =2$ imposed by the Planck
scalar spectrum bounds \protect\cite{Planck}. The yellow and blue ellipses
represent the expected constraint on the parameter space with $1\protect%
\sigma $ and $2\protect\sigma $ significance, respectively.}
\label{Contour1}
\end{figure}

We can now compare our predictions for $\gamma =2$ to the observational
results found in \cite{Planck}. In the case of the scalar power spectrum,
for the form written in eq.~(\ref{PS1}), in this reference, it was found
that for $k_{0}=0.05Mpc^{-1}$: 
\begin{align}
A_{\zeta }^{2}& =\left( 2.196_{-0.060}^{+0.051}\right) \times 10^{-9},
\label{Azeta} \\
n_{\text{s}}-1& =-0.0371\pm 0.0057.  \label{Ns1}
\end{align}%
These values constrain the set of parameters $(\beta ,p_{0},\lambda )$ of
our model. It is convenient to study the case where $p_{0}$ is fixed due to
the fact that $p_{0}$ always appears as $\ln (p_{0})$ in eqs. (\ref{PSt})
and (\ref{SpecIn2}), and then these observational constraints will predict a
big uncertainty for this parameter. If we do this, then the observations at $%
k_{0}$ constrain the parameters $\beta $ and $\lambda $. For the case of $%
p_{0}=10^{135}$, the results are shown in Fig.~\ref{Contour1}.

Fig.~\ref{Contour1} shows the $(\beta ,\lambda )$ joint error bounds. The
dot in the plot shows the set $(\bar{\beta}=1.3353,\bar{\lambda}%
=1.3013\times 10^{4})$ that maximises the likelihood, i.e.~that give the
observational central values in eqs.~(\ref{Azeta})-(\ref{Ns1}), when
replaced in the theoretical equations (\ref{PSt}) and (\ref{SpecIn2}), for $%
p_{0}=10^{135}$. Using the Fisher matrix technique, we find the approximated
regions with $1\sigma $ and $2\sigma $ joint errors, represented by the
yellow and blue regions, respectively. We observe that both parameters are
highly correlated, in fact $Corr(\lambda ,\beta )=99.7\%$. The errors found
yield $\lambda =(1.3013\pm 0.2020)\times 10^{4}$ and $\beta =1.3353\pm 0.2108
$.

\begin{figure}[h]
\centering
\scalebox{0.43}{\includegraphics{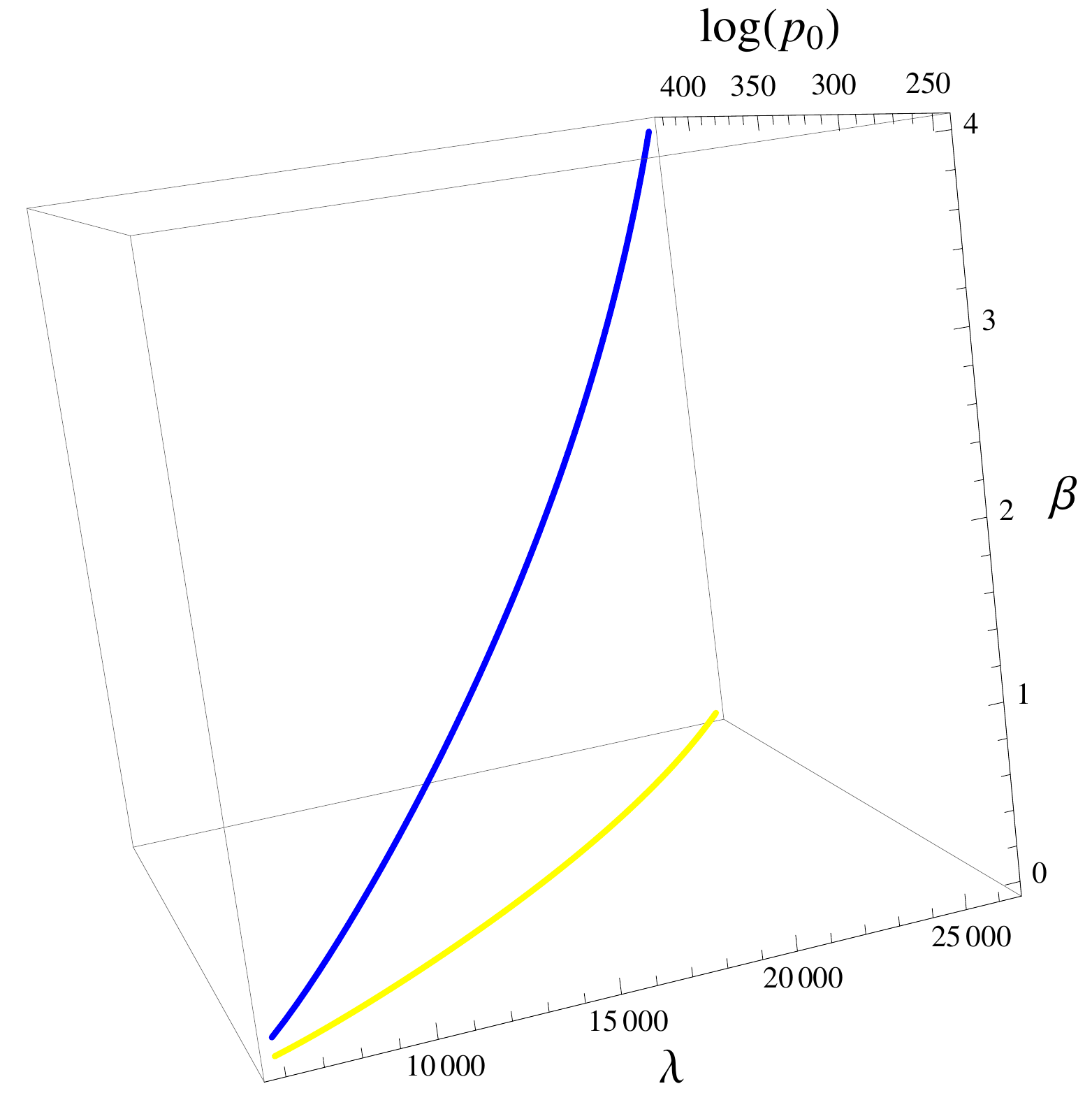}}
\caption{The set of values for $(\protect\beta ,\protect\lambda ,\log
(p_{0}))$ satisfying the experimental constraints of $A_{\protect\zeta }^{2}$%
, and $n_{\text{s}}$ at $k_{0}=0.05Mpc^{-1}$, represented by the blue line.
We also show the projection of this line onto the $(\protect\lambda ,\log
(p_{0}))$ plane, represented by the yellow line.}
\label{Const1}
\end{figure}

Since our model predicts a $k$-dependent spectral index at next order, we
could improve our approximation of the power spectrum by expanding $\mathcal{%
P}_{\zeta }$ as: 
\begin{equation}
\mathcal{P}_{\zeta }(k)\sim A_{\zeta }^{2}\left( \frac{k}{k_{0}}\right) ^{n_{%
\text{s}}(k_{0})-1+\frac{1}{2}\ln (k/k_{0})\alpha +\frac{1}{6}\ln
^{2}(k/k_{0})\alpha _{2}},
\end{equation}%
where, at some pivotal scale $k_{0}$, the index $n_{\text{s}}$ is given by
eq.~(\ref{SpecIn2}), while $\alpha $ and $\alpha _{2}$ are given by: 
\begin{align}
& \alpha \equiv \left. \frac{dn_{\text{s}}}{d\ln (k)}\right\vert _{k_{0}}=%
\frac{4\beta }{\left( E+2\ln (\tilde{\lambda}k_{0})\right) ^{2}},
\label{AlphaT} \\
& \alpha _{2}\equiv \left. \frac{d^{2}n_{\text{s}}}{d\ln k^{2}}\right\vert
_{k_{0}}=\frac{-16\beta }{\left( E+2\ln (\tilde{\lambda}k_{0})\right) ^{3}}.
\label{Alpha2T}
\end{align}%
In \cite{Planck} it was found that, for $k_{0}=0.05Mpc^{-1}$, the
constraints are: 
\begin{align}
n_{\text{s}}-1& =-0.0432_{-0.0063}^{+0.0068},  \label{Ns} \\
\alpha & =0.000_{-0.013}^{+0.016},  \label{Alpha} \\
\alpha _{2}& =0.017_{-0.014}^{+0.016},  \label{Alpha2}
\end{align}%
while $A_{\zeta }^{2}$ is given by eq.~(\ref{Azeta}). Again, these values
constrain the set of parameters $(\beta ,p_{0},\lambda )$ of our model. Fig.~%
\ref{Const1} shows a sample of the set of parameters that give the central
observational values of (\ref{Azeta}) and (\ref{Ns}).

In Fig.~\ref{Const1} we draw a 3-dimensional box with axes $(\beta ,\log
(p_{0}),\lambda )$. The blue line shows the concordant set of parameters
while the yellow line is its projection onto the plane $(0,\log (p_{0}),\beta
) $. Here, we only plot a sample range of parameters such that $\beta $ is
in the range of $0.1\leqslant \beta \leqslant 4$. 
% From this plot, we can see that, for instance, for $\beta=1$, the most likely values are $\lambda\approx 7900$ and $p_0\approx 10^{281}$. 

\begin{figure}[h]
\begin{center}
\scalebox{0.5}{\includegraphics{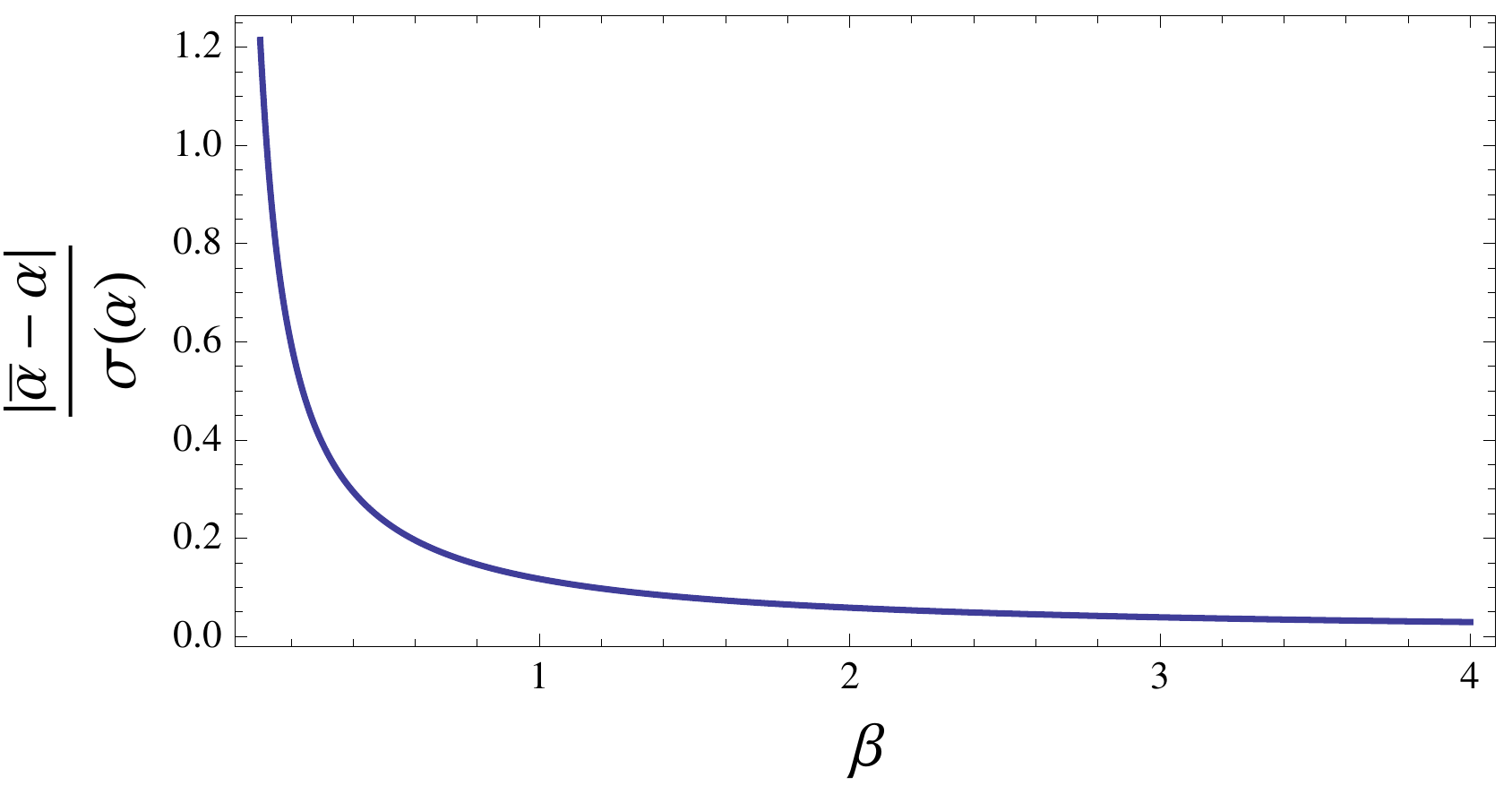}}
\end{center}
\caption{The deviation of $\protect\alpha $ (in terms of the error $\protect%
\sigma (\protect\alpha )$) from the central experimental value of $\protect%
\alpha $ from ref. \protect\cite{Planck} as a function of $\protect\beta $.}
\label{Sig}
\end{figure}

Not all  of parameter values in the set shown in Fig.~\ref{Const1} satisfy
the observational central values found for $\alpha $ and $\alpha _{2}$ in
ref.\cite{Planck}. In the case of $\alpha $, we can calculate the predicted
values by using the set of parameters in Fig.~\ref{Const1} and eq.~(\ref%
{AlphaT}). This set of predicted values for $\alpha $ will always have
positive values and therefore they will have some deviation from the central
observed value $\bar{\alpha}=0$. Fig.~\ref{Sig} shows this deviation (in
terms of the error $\sigma (\alpha )=0.016$) as a function of the parameter $%
\beta $. In this figure we see that the deviation is less than $1.2\sigma
(\alpha )$ for the sampled values of $\beta $. 

\begin{figure}[h]
\begin{center}
\scalebox{0.53}{\includegraphics{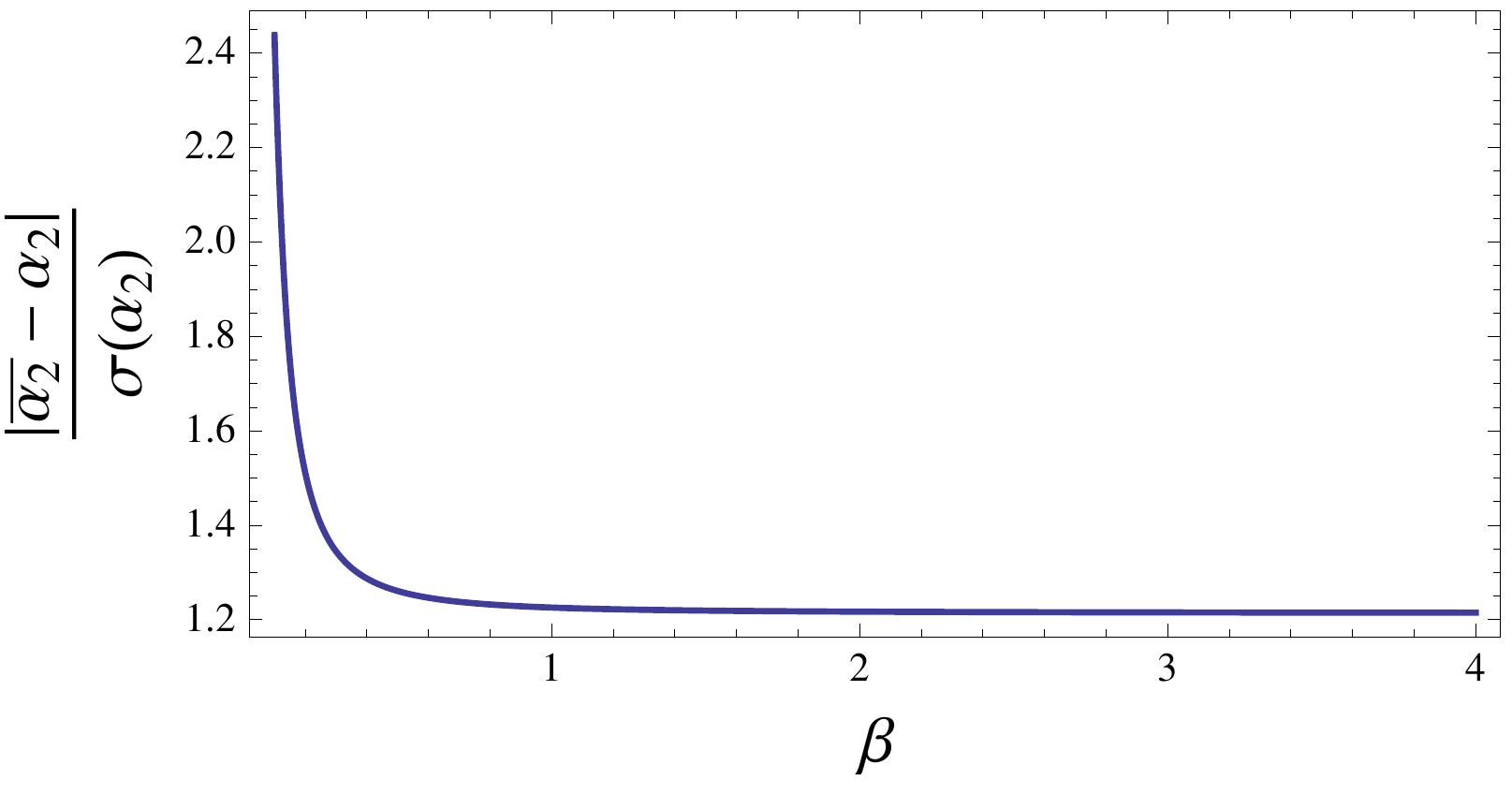}}
\end{center}
\caption{The deviation of $\protect\alpha _{2}$ (in terms of the error $%
\protect\sigma (\protect\alpha _{2})$) from the central observational value
of $\protect\alpha _{2}$ as a function of $\protect\beta $.}
\label{Sig2}
\end{figure}

On the other hand, in the case of $\alpha _{2}$ we notice that,
observationally, it is expected to be positive at a level of about $%
1.2\sigma (\alpha _{2})$ (see eq.~(\ref{Alpha2})), where $\sigma (\alpha
_{2})=0.014$. However, our theory  always predicts a negative value, as eq.~(%
\ref{Alpha2T}) shows \footnote{%
In principle, $\alpha _{2}$ could be positive if the term $E+2\ln (\tilde{%
\lambda}k_{0})$ were negative, but this term is always positive if we want
to satisfy eq.~(\ref{Ns}).}. If now we do the same study we just did for $%
\alpha $, we will always predict  values with a deviation of more than $%
1\sigma (\alpha _{2})$, as shown in Fig.~\ref{Sig2}. From this plot we can
see that the deviation of $\alpha _{2}$ is less than $2.5\sigma $ for the
set of sampled $\beta $ values.

Therefore, even though there are predicted deviations from the central
values of $\alpha $ and $\alpha _{2}$, these do not rule out the model
because they are sufficiently small.

On the other hand, we could also find the appropriate value for the
parameter $b$ which sets the tensor-to-scalar ratio. In \cite{Planck}  a
maximum value of 
\begin{equation}
r<0.12
\end{equation}%
was found at the pivotal scale $k_{0}=0.002Mpc^{-1}$. The corresponding
condition on the parameter $b$ will depend on the values we take for $(D,%
\tilde{\lambda},\beta )$ (or, equivalently for the set $(p_{0},\lambda
,\beta )$). For the case with a fixed $p_{0}=10^{135}$, and the values $(%
\bar{\lambda},\bar{\beta})$, we must have $b>31.3$ in order to satisfy the
lower limit on $r$. This means that in the high-energy regime, the speed of
gravity exceeds the speed of light. We could also contemplate scenarios
where the parameter $\beta $ is different for the scalar and tensor case, so
we would have separate $\beta _{\text{s}}$ and $\beta _{\text{T}}$
parameters with different values.

\section{Conclusions}

We have studied the cosmological implications of a specific class of MDRs,
when they are combined with general relativity to create an early Universe
cosmology. The resulting cosmology is equivalent to a dual model displaying
intermediate inflation driven by modified gravity, in the guise of rainbow
gravity. The specific case $\gamma =2$ is very interesting both from the
quantum gravitational and the cosmological perspectives. When $\beta =0$ the
MDRs are known to model the running from four dimensions in the IR to two
dimensions in the UV (\cite{dimred}), as suggested by numerous quantum
gravity studies (\cite%
{Ambjorn:1997jf,Horava:2009if,Ambjorn:2005db,Lauscher:2005qz,Kommu:2011wd,Giasemidis:2012rf,Alesci:2011cg,Modesto:2008jz,Calcagni:2010pa}%
). The associated cosmological model then predicts exact scale-invariance
for the primordial fluctuations. If we resort to fractional powers, with $%
\gamma $ slightly larger than 2, it is possible, but contrived, to produce
departures from exact scale-invariance. Keeping $\gamma =2$, but building in
a very long UV transient (as suggested in \cite{dimred}) , is a more contrived possibility. The MDRs
associated with intermediate inflation resolve this problem. Using both the
spectral and Hausdorff dimension, we found that when $\gamma =2$ but $\beta
\neq 0$, after an apparent running from 4 to 2 dimensions at the scale $%
\lambda $, the dimension never settles at 2, but drifts upwards very slowly
over several orders of magnitude in energy. This can be achieved with
natural values for $\beta $, say 1 or 2. This phenomenon would be hard to
measure in simulations (e.g.~in Causal Dynamical Triangulations, abbreviated as CDTs) without the development of specific
methods. And yet the logarithmic corrections in the MDRs are typical of
renormalisation group corrections. From the observational point of view this
is precisely what supplies a natural mechanism for inducing departures from
scale invariance in the primordial power spectrum, as we explicitly
confirmed with calculations of the expected perturbations.

We studied  the cosmological observational implications of this model in
detail. We found that the model is able to predict primordial fluctuations
in accordance with observations, in an expanding (but not inflating) early
Universe dominated by radiation. Specifically, setting $\gamma =2$ and $%
P/\rho \equiv w=1/3$, we showed that a considerable range of the remaining
model parameters can lead to viable predictions for the resulting
cosmological density perturbations. We find that $\beta $ does not need to
be highly contrived, since it can be at least in the range 1-5, while the
length scale $\lambda $ must be four orders of magnitude above the Planck
scale in order to obtain $n_{S}\sim 0.96$ and explain the amplitude $%
A_{\zeta }$. In addition, we introduced a new parameter in the theory to
model tensor perturbations, which is capable of explaining the observed
upper bound of the tensor-to-scalar ratio. The parameter choices made in
this model can be compared with those typically made to arrive at best-fit
inflationary models \cite{Planck}.

Our model makes clear prediction for the running of the spectral index,
encoded in eq.~(\ref{AlphaT}) and (\ref{Alpha2T}). These may go against
current observational \textquotedblleft trends\textquotedblright\ (e.g.~in
terms of sign for one of these parameters), but typically at only around $%
1\sigma $ (and never by more than $2.5\sigma $ in the range of parameters
considered). This explains our label \textquotedblleft
trends\textquotedblright\ for these observations, which are not
discriminating enough for the purpose of the model proposed here. However,
this very remark leaves open the prospect of falsifying or verifying these
models, as the data sharpens and becomes more discriminating. The running of
the spectral index could be the ideal test for this class of models.

On the other hand, predictions based on non-Gaussianity (e.g.~$f_{NL}$)
remain to be worked out. We emphasise that we cannot simply read off the
results obtained for single field intermediate inflation (\cite{Chen:2010xka}%
). A rather non-trivial fresh calculation is needed, since we would have to
calculate the third-order action for higher-order derivative theories. This
will be the subject of a future paper.

We close by noting that the models considered here are among the simplest of
those leading to rainbow intermediate inflation. By fixing $\gamma =2$ we
are requiring that in the Einstein frame the Universe be filled with
radiation, i.e.~$w=1/3$ (cf. eq.~(27) in \cite{inter-rainb}). In the rainbow
frame, a case of interest is $\beta =1$, where we have that the scale factor
evolves as: 
\begin{equation}
a\left( \tilde{t}\right) \propto \lambda k\exp {\left( \frac{a_{0}\tilde{t}%
^{1/2}}{\lambda k}\right) }\;,
\end{equation}%
where here we have used some notation defined in \cite{inter-rainb}. These
models softly break the conformal invariance of the gravitational coupling
of fluctuations, which is peculiar to these models when $\beta =0$ as it was shown in \cite{rainbowred}. This is the root
of their phenomenological and theoretical interest as alternatives to
inflation for a detailed explanation of the structure observed in the cosmic
microwave background radiation.

\section*{Acknowledgments}

We thank G. Amelino-Camelia, M. Arzano, A. Eichhorn, G. Calcagni and G. Gubitosi for the many
discussions that led to this project. JM acknowledges STFC consolidated
grant support. ML was funded by Becas Chile.

\bibliographystyle{apsrev4-1}
\bibliography{refs-macarena}

\begin{thebibliography}{44}%
\makeatletter
\providecommand \@ifxundefined [1]{%
 \@ifx{#1\undefined}
}%
\providecommand \@ifnum [1]{%
 \ifnum #1\expandafter \@firstoftwo
 \else \expandafter \@secondoftwo
 \fi
}%
\providecommand \@ifx [1]{%
 \ifx #1\expandafter \@firstoftwo
 \else \expandafter \@secondoftwo
 \fi
}%
\providecommand \natexlab [1]{#1}%
\providecommand \enquote  [1]{``#1''}%
\providecommand \bibnamefont  [1]{#1}%
\providecommand \bibfnamefont [1]{#1}%
\providecommand \citenamefont [1]{#1}%
\providecommand \href@noop [0]{\@secondoftwo}%
\providecommand \href [0]{\begingroup \@sanitize@url \@href}%
\providecommand \@href[1]{\@@startlink{#1}\@@href}%
\providecommand \@@href[1]{\endgroup#1\@@endlink}%
\providecommand \@sanitize@url [0]{\catcode `\\12\catcode `\$12\catcode
  `\&12\catcode `\#12\catcode `\^12\catcode `\_12\catcode `\%12\relax}%
\providecommand \@@startlink[1]{}%
\providecommand \@@endlink[0]{}%
\providecommand \url  [0]{\begingroup\@sanitize@url \@url }%
\providecommand \@url [1]{\endgroup\@href {#1}{\urlprefix }}%
\providecommand \urlprefix  [0]{URL }%
\providecommand \Eprint [0]{\href }%
\providecommand \doibase [0]{http://dx.doi.org/}%
\providecommand \selectlanguage [0]{\@gobble}%
\providecommand \bibinfo  [0]{\@secondoftwo}%
\providecommand \bibfield  [0]{\@secondoftwo}%
\providecommand \translation [1]{[#1]}%
\providecommand \BibitemOpen [0]{}%
\providecommand \bibitemStop [0]{}%
\providecommand \bibitemNoStop [0]{.\EOS\space}%
\providecommand \EOS [0]{\spacefactor3000\relax}%
\providecommand \BibitemShut  [1]{\csname bibitem#1\endcsname}%
\let\auto@bib@innerbib\@empty
%</preamble>
\bibitem [{\citenamefont {Ambjorn}\ \emph
  {et~al.}(2005{\natexlab{a}})\citenamefont {Ambjorn}, \citenamefont
  {Jurkiewicz},\ and\ \citenamefont {Loll}}]{Lollprl}%
  \BibitemOpen
  \bibfield  {author} {\bibinfo {author} {\bibfnamefont {J.}~\bibnamefont
  {Ambjorn}}, \bibinfo {author} {\bibfnamefont {J.}~\bibnamefont {Jurkiewicz}},
  \ and\ \bibinfo {author} {\bibfnamefont {R.}~\bibnamefont {Loll}},\ }\href
  {\doibase 10.1103/PhysRevLett.95.171301} {\bibfield  {journal} {\bibinfo
  {journal} {Phys.Rev.Lett.}\ }\textbf {\bibinfo {volume} {95}},\ \bibinfo
  {pages} {171301} (\bibinfo {year} {2005}{\natexlab{a}})},\ \Eprint
  {http://arxiv.org/abs/hep-th/0505113} {arXiv:hep-th/0505113 [hep-th]}
  \BibitemShut {NoStop}%
%%CITATION = HEP-TH/0505113;%%
\bibitem [{\citenamefont {Litim}(2004)}]{Litim}%
  \BibitemOpen
  \bibfield  {author} {\bibinfo {author} {\bibfnamefont {D.~F.}\ \bibnamefont
  {Litim}},\ }\href {\doibase 10.1103/PhysRevLett.92.201301} {\bibfield
  {journal} {\bibinfo  {journal} {Phys.Rev.Lett.}\ }\textbf {\bibinfo {volume}
  {92}},\ \bibinfo {pages} {201301} (\bibinfo {year} {2004})},\ \Eprint
  {http://arxiv.org/abs/hep-th/0312114} {arXiv:hep-th/0312114 [hep-th]}
  \BibitemShut {NoStop}%
%%CITATION = HEP-TH/0312114;%%
\bibitem [{\citenamefont {Lauscher}\ and\ \citenamefont
  {Reuter}(2005{\natexlab{a}})}]{Reuter}%
  \BibitemOpen
  \bibfield  {author} {\bibinfo {author} {\bibfnamefont {O.}~\bibnamefont
  {Lauscher}}\ and\ \bibinfo {author} {\bibfnamefont {M.}~\bibnamefont
  {Reuter}},\ }\href {\doibase 10.1088/1126-6708/2005/10/050} {\bibfield
  {journal} {\bibinfo  {journal} {JHEP}\ }\textbf {\bibinfo {volume} {0510}},\
  \bibinfo {pages} {050} (\bibinfo {year} {2005}{\natexlab{a}})},\ \Eprint
  {http://arxiv.org/abs/hep-th/0508202} {arXiv:hep-th/0508202 [hep-th]}
  \BibitemShut {NoStop}%
%%CITATION = HEP-TH/0508202;%%
\bibitem [{\citenamefont {Horava}(2009{\natexlab{a}})}]{Hl}%
  \BibitemOpen
  \bibfield  {author} {\bibinfo {author} {\bibfnamefont {P.}~\bibnamefont
  {Horava}},\ }\href {\doibase 10.1103/PhysRevD.79.084008} {\bibfield
  {journal} {\bibinfo  {journal} {Phys.Rev.}\ }\textbf {\bibinfo {volume}
  {D79}},\ \bibinfo {pages} {084008} (\bibinfo {year} {2009}{\natexlab{a}})},\
  \Eprint {http://arxiv.org/abs/0901.3775} {arXiv:0901.3775 [hep-th]}
  \BibitemShut {NoStop}%
%%CITATION = ARXIV:0901.3775;%%
\bibitem [{\citenamefont {Horava}(2009{\natexlab{b}})}]{HLspec}%
  \BibitemOpen
  \bibfield  {author} {\bibinfo {author} {\bibfnamefont {P.}~\bibnamefont
  {Horava}},\ }\href {\doibase 10.1103/PhysRevLett.102.161301} {\bibfield
  {journal} {\bibinfo  {journal} {Phys.Rev.Lett.}\ }\textbf {\bibinfo {volume}
  {102}},\ \bibinfo {pages} {161301} (\bibinfo {year} {2009}{\natexlab{b}})},\
  \Eprint {http://arxiv.org/abs/0902.3657} {arXiv:0902.3657 [hep-th]}
  \BibitemShut {NoStop}%
%%CITATION = ARXIV:0902.3657;%%
\bibitem [{\citenamefont {Alesci}\ and\ \citenamefont
  {Arzano}(2012)}]{Alesci:2011cg}%
  \BibitemOpen
  \bibfield  {author} {\bibinfo {author} {\bibfnamefont {E.}~\bibnamefont
  {Alesci}}\ and\ \bibinfo {author} {\bibfnamefont {M.}~\bibnamefont
  {Arzano}},\ }\href {\doibase 10.1016/j.physletb.2011.12.026} {\bibfield
  {journal} {\bibinfo  {journal} {Phys.Lett.}\ }\textbf {\bibinfo {volume}
  {B707}},\ \bibinfo {pages} {272} (\bibinfo {year} {2012})},\ \Eprint
  {http://arxiv.org/abs/1108.1507} {arXiv:1108.1507 [gr-qc]} \BibitemShut
  {NoStop}%
%%CITATION = ARXIV:1108.1507;%%
\bibitem [{\citenamefont {Benedetti}(2009)}]{Benedetti:2008gu}%
  \BibitemOpen
  \bibfield  {author} {\bibinfo {author} {\bibfnamefont {D.}~\bibnamefont
  {Benedetti}},\ }\href {\doibase 10.1103/PhysRevLett.102.111303} {\bibfield
  {journal} {\bibinfo  {journal} {Phys.Rev.Lett.}\ }\textbf {\bibinfo {volume}
  {102}},\ \bibinfo {pages} {111303} (\bibinfo {year} {2009})},\ \Eprint
  {http://arxiv.org/abs/0811.1396} {arXiv:0811.1396 [hep-th]} \BibitemShut
  {NoStop}%
%%CITATION = ARXIV:0811.1396;%%
\bibitem [{\citenamefont {Modesto}(2009{\natexlab{a}})}]{Modesto1}%
  \BibitemOpen
  \bibfield  {author} {\bibinfo {author} {\bibfnamefont {L.}~\bibnamefont
  {Modesto}},\ }\href {\doibase 10.1088/0264-9381/26/24/242002} {\bibfield
  {journal} {\bibinfo  {journal} {Class.Quant.Grav.}\ }\textbf {\bibinfo
  {volume} {26}},\ \bibinfo {pages} {242002} (\bibinfo {year}
  {2009}{\natexlab{a}})},\ \Eprint {http://arxiv.org/abs/0812.2214}
  {arXiv:0812.2214 [gr-qc]} \BibitemShut {NoStop}%
%%CITATION = ARXIV:0812.2214;%%
\bibitem [{\citenamefont {Caravelli}\ and\ \citenamefont
  {Modesto}(2009)}]{Caravelli}%
  \BibitemOpen
  \bibfield  {author} {\bibinfo {author} {\bibfnamefont {F.}~\bibnamefont
  {Caravelli}}\ and\ \bibinfo {author} {\bibfnamefont {L.}~\bibnamefont
  {Modesto}},\ }\href@noop {} {\  (\bibinfo {year} {2009})},\ \Eprint
  {http://arxiv.org/abs/0905.2170} {arXiv:0905.2170 [gr-qc]} \BibitemShut
  {NoStop}%
%%CITATION = ARXIV:0905.2170;%%
\bibitem [{\citenamefont {Magliaro}\ \emph {et~al.}(2009)\citenamefont
  {Magliaro}, \citenamefont {Perini},\ and\ \citenamefont
  {Modesto}}]{Magliaro}%
  \BibitemOpen
  \bibfield  {author} {\bibinfo {author} {\bibfnamefont {E.}~\bibnamefont
  {Magliaro}}, \bibinfo {author} {\bibfnamefont {C.}~\bibnamefont {Perini}}, \
  and\ \bibinfo {author} {\bibfnamefont {L.}~\bibnamefont {Modesto}},\
  }\href@noop {} {\  (\bibinfo {year} {2009})},\ \Eprint
  {http://arxiv.org/abs/0911.0437} {arXiv:0911.0437 [gr-qc]} \BibitemShut
  {NoStop}%
%%CITATION = ARXIV:0911.0437;%%
\bibitem [{\citenamefont {Modesto}\ and\ \citenamefont
  {Nicolini}(2010)}]{Modesto2}%
  \BibitemOpen
  \bibfield  {author} {\bibinfo {author} {\bibfnamefont {L.}~\bibnamefont
  {Modesto}}\ and\ \bibinfo {author} {\bibfnamefont {P.}~\bibnamefont
  {Nicolini}},\ }\href {\doibase 10.1103/PhysRevD.81.104040} {\bibfield
  {journal} {\bibinfo  {journal} {Phys.Rev.}\ }\textbf {\bibinfo {volume}
  {D81}},\ \bibinfo {pages} {104040} (\bibinfo {year} {2010})},\ \Eprint
  {http://arxiv.org/abs/0912.0220} {arXiv:0912.0220 [hep-th]} \BibitemShut
  {NoStop}%
%%CITATION = ARXIV:0912.0220;%%
\bibitem [{\citenamefont {Calcagni}(2011{\natexlab{a}})}]{Calcagni1}%
  \BibitemOpen
  \bibfield  {author} {\bibinfo {author} {\bibfnamefont {G.}~\bibnamefont
  {Calcagni}},\ }\href {\doibase 10.1016/j.physletb.2011.01.063} {\bibfield
  {journal} {\bibinfo  {journal} {Phys.Lett.}\ }\textbf {\bibinfo {volume}
  {B697}},\ \bibinfo {pages} {251} (\bibinfo {year} {2011}{\natexlab{a}})},\
  \Eprint {http://arxiv.org/abs/1012.1244} {arXiv:1012.1244 [hep-th]}
  \BibitemShut {NoStop}%
%%CITATION = ARXIV:1012.1244;%%
\bibitem [{\citenamefont {Calcagni}(2013)}]{Calcagni2}%
  \BibitemOpen
  \bibfield  {author} {\bibinfo {author} {\bibfnamefont {G.}~\bibnamefont
  {Calcagni}},\ }\href {\doibase 10.1103/PhysRevE.87.012123} {\bibfield
  {journal} {\bibinfo  {journal} {Phys.Rev.}\ }\textbf {\bibinfo {volume}
  {E87}},\ \bibinfo {pages} {012123} (\bibinfo {year} {2013})},\ \Eprint
  {http://arxiv.org/abs/1205.5046} {arXiv:1205.5046 [hep-th]} \BibitemShut
  {NoStop}%
%%CITATION = ARXIV:1205.5046;%%
\bibitem [{\citenamefont {Sotiriou}\ \emph {et~al.}(2011)\citenamefont
  {Sotiriou}, \citenamefont {Visser},\ and\ \citenamefont
  {Weinfurtner}}]{visser}%
  \BibitemOpen
  \bibfield  {author} {\bibinfo {author} {\bibfnamefont {T.~P.}\ \bibnamefont
  {Sotiriou}}, \bibinfo {author} {\bibfnamefont {M.}~\bibnamefont {Visser}}, \
  and\ \bibinfo {author} {\bibfnamefont {S.}~\bibnamefont {Weinfurtner}},\
  }\href {\doibase 10.1103/PhysRevD.84.104018} {\bibfield  {journal} {\bibinfo
  {journal} {Phys.Rev.}\ }\textbf {\bibinfo {volume} {D84}},\ \bibinfo {pages}
  {104018} (\bibinfo {year} {2011})},\ \Eprint {http://arxiv.org/abs/1105.6098}
  {arXiv:1105.6098 [hep-th]} \BibitemShut {NoStop}%
%%CITATION = ARXIV:1105.6098;%%
\bibitem [{\citenamefont {Amelino-Camelia}\ \emph
  {et~al.}(2013{\natexlab{a}})\citenamefont {Amelino-Camelia}, \citenamefont
  {Arzano}, \citenamefont {Gubitosi},\ and\ \citenamefont {Magueijo}}]{dsrrsd}%
  \BibitemOpen
  \bibfield  {author} {\bibinfo {author} {\bibfnamefont {G.}~\bibnamefont
  {Amelino-Camelia}}, \bibinfo {author} {\bibfnamefont {M.}~\bibnamefont
  {Arzano}}, \bibinfo {author} {\bibfnamefont {G.}~\bibnamefont {Gubitosi}}, \
  and\ \bibinfo {author} {\bibfnamefont {J.}~\bibnamefont {Magueijo}},\
  }\href@noop {} {\  (\bibinfo {year} {2013}{\natexlab{a}})},\ \Eprint
  {http://arxiv.org/abs/1311.3135} {arXiv:1311.3135 [gr-qc]} \BibitemShut
  {NoStop}%
%%CITATION = ARXIV:1311.3135;%%
\bibitem [{\citenamefont {Amelino-Camelia}\ \emph
  {et~al.}(2013{\natexlab{b}})\citenamefont {Amelino-Camelia}, \citenamefont
  {Arzano}, \citenamefont {Gubitosi},\ and\ \citenamefont {Magueijo}}]{dimred}%
  \BibitemOpen
  \bibfield  {author} {\bibinfo {author} {\bibfnamefont {G.}~\bibnamefont
  {Amelino-Camelia}}, \bibinfo {author} {\bibfnamefont {M.}~\bibnamefont
  {Arzano}}, \bibinfo {author} {\bibfnamefont {G.}~\bibnamefont {Gubitosi}}, \
  and\ \bibinfo {author} {\bibfnamefont {J.}~\bibnamefont {Magueijo}},\ }\href
  {\doibase 10.1103/PhysRevD.87.123532} {\bibfield  {journal} {\bibinfo
  {journal} {Phys. Rev. D}\ }\textbf {\bibinfo {volume} {87}},\ \bibinfo
  {pages} {123532} (\bibinfo {year} {2013}{\natexlab{b}})},\ \Eprint
  {http://arxiv.org/abs/1305.3153} {1305.3153} \BibitemShut {NoStop}%
\bibitem [{\citenamefont {Magueijo}\ and\ \citenamefont
  {Smolin}(2004)}]{rainbowDSR}%
  \BibitemOpen
  \bibfield  {author} {\bibinfo {author} {\bibfnamefont {J.}~\bibnamefont
  {Magueijo}}\ and\ \bibinfo {author} {\bibfnamefont {L.}~\bibnamefont
  {Smolin}},\ }\href {\doibase 10.1088/0264-9381/21/7/001} {\bibfield
  {journal} {\bibinfo  {journal} {Class. Quant. Grav.}\ }\textbf {\bibinfo
  {volume} {21}},\ \bibinfo {pages} {1725} (\bibinfo {year} {2004})},\ \Eprint
  {http://arxiv.org/abs/gr-qc/0305055} {arXiv:gr-qc/0305055} \BibitemShut
  {NoStop}%
\bibitem [{\citenamefont {Amelino-Camelia}\ \emph
  {et~al.}(2013{\natexlab{c}})\citenamefont {Amelino-Camelia}, \citenamefont
  {Arzano}, \citenamefont {Gubitosi},\ and\ \citenamefont
  {Magueijo}}]{rainbowred}%
  \BibitemOpen
  \bibfield  {author} {\bibinfo {author} {\bibfnamefont {G.}~\bibnamefont
  {Amelino-Camelia}}, \bibinfo {author} {\bibfnamefont {M.}~\bibnamefont
  {Arzano}}, \bibinfo {author} {\bibfnamefont {G.}~\bibnamefont {Gubitosi}}, \
  and\ \bibinfo {author} {\bibfnamefont {J.}~\bibnamefont {Magueijo}},\ }\href
  {\doibase 10.1103/PhysRevD.88.041303} {\bibfield  {journal} {\bibinfo
  {journal} {Phys. Rev. D}\ }\textbf {\bibinfo {volume} {88}},\ \bibinfo
  {pages} {041303} (\bibinfo {year} {2013}{\natexlab{c}})},\ \Eprint
  {http://arxiv.org/abs/1307.0745} {1307.0745} \BibitemShut {NoStop}%
\bibitem [{\citenamefont {Magueijo}(2008{\natexlab{a}})}]{csdot}%
  \BibitemOpen
  \bibfield  {author} {\bibinfo {author} {\bibfnamefont {J.}~\bibnamefont
  {Magueijo}},\ }\href {\doibase 10.1103/PhysRevLett.100.231302} {\bibfield
  {journal} {\bibinfo  {journal} {Phys.Rev.Lett.}\ }\textbf {\bibinfo {volume}
  {100}},\ \bibinfo {pages} {231302} (\bibinfo {year} {2008}{\natexlab{a}})},\
  \Eprint {http://arxiv.org/abs/arXiv:0803.0859} {arXiv:arXiv:0803.0859
  [astro-ph]} \BibitemShut {NoStop}%
%%CITATION = ARXIV:0803.0859;%%
\bibitem [{\citenamefont {Magueijo}(2009)}]{bim}%
  \BibitemOpen
  \bibfield  {author} {\bibinfo {author} {\bibfnamefont {J.}~\bibnamefont
  {Magueijo}},\ }\href {\doibase 10.1103/PhysRevD.79.043525} {\bibfield
  {journal} {\bibinfo  {journal} {Phys.Rev.}\ }\textbf {\bibinfo {volume}
  {D79}},\ \bibinfo {pages} {043525} (\bibinfo {year} {2009})},\ \Eprint
  {http://arxiv.org/abs/0807.1689} {arXiv:0807.1689 [gr-qc]} \BibitemShut
  {NoStop}%
%%CITATION = ARXIV:0807.1689;%%
\bibitem [{\citenamefont {Magueijo}(2008{\natexlab{b}})}]{Mag}%
  \BibitemOpen
  \bibfield  {author} {\bibinfo {author} {\bibfnamefont {J.}~\bibnamefont
  {Magueijo}},\ }\href {\doibase 10.1088/0264-9381/25/20/202002} {\bibfield
  {journal} {\bibinfo  {journal} {Class.Quant.Grav.}\ }\textbf {\bibinfo
  {volume} {25}},\ \bibinfo {pages} {202002} (\bibinfo {year}
  {2008}{\natexlab{b}})},\ \Eprint {http://arxiv.org/abs/0807.1854}
  {arXiv:0807.1854 [gr-qc]} \BibitemShut {NoStop}%
%%CITATION = ARXIV:0807.1854;%%
\bibitem [{\citenamefont {Garattini}\ and\ \citenamefont
  {Sakellariadou}(2012)}]{Garattini}%
  \BibitemOpen
  \bibfield  {author} {\bibinfo {author} {\bibfnamefont {R.}~\bibnamefont
  {Garattini}}\ and\ \bibinfo {author} {\bibfnamefont {M.}~\bibnamefont
  {Sakellariadou}},\ }\href@noop {} {\  (\bibinfo {year} {2012})},\ \Eprint
  {http://arxiv.org/abs/1212.4987} {arXiv:1212.4987 [gr-qc]} \BibitemShut
  {NoStop}%
%%CITATION = ARXIV:1212.4987;%%
\bibitem [{\citenamefont {Starobinsky}(2005)}]{Starob}%
  \BibitemOpen
  \bibfield  {author} {\bibinfo {author} {\bibfnamefont {A.~A.}\ \bibnamefont
  {Starobinsky}},\ }\href {\doibase 10.1134/1.2121807} {\bibfield  {journal}
  {\bibinfo  {journal} {JETP Lett.}\ }\textbf {\bibinfo {volume} {82}},\
  \bibinfo {pages} {169} (\bibinfo {year} {2005})},\ \Eprint
  {http://arxiv.org/abs/astro-ph/0507193} {arXiv:astro-ph/0507193 [astro-ph]}
  \BibitemShut {NoStop}%
%%CITATION = ASTRO-PH/0507193;%%
\bibitem [{\citenamefont {Barrow}(1990)}]{intinfl0}%
  \BibitemOpen
  \bibfield  {author} {\bibinfo {author} {\bibfnamefont {J.~D.}\ \bibnamefont
  {Barrow}},\ }\href {\doibase 10.1016/0370-2693(90)91007-X} {\bibfield
  {journal} {\bibinfo  {journal} {Phys.Lett.}\ }\textbf {\bibinfo {volume}
  {B235}},\ \bibinfo {pages} {40} (\bibinfo {year} {1990})}\BibitemShut
  {NoStop}%
%%CITATION = PHLTA,B249,406;%%
\bibitem [{\citenamefont {Barrow}\ and\ \citenamefont
  {Saich}(1990)}]{intinfl1}%
  \BibitemOpen
  \bibfield  {author} {\bibinfo {author} {\bibfnamefont {J.~D.}\ \bibnamefont
  {Barrow}}\ and\ \bibinfo {author} {\bibfnamefont {P.}~\bibnamefont {Saich}},\
  }\href {\doibase 10.1016/0370-2693(90)91007-X} {\bibfield  {journal}
  {\bibinfo  {journal} {Phys.Lett.}\ }\textbf {\bibinfo {volume} {B249}},\
  \bibinfo {pages} {406} (\bibinfo {year} {1990})}\BibitemShut {NoStop}%
%%CITATION = PHLTA,B249,406;%%
\bibitem [{\citenamefont {Barrow}\ and\ \citenamefont
  {Liddle}(1993)}]{intinfl2}%
  \BibitemOpen
  \bibfield  {author} {\bibinfo {author} {\bibfnamefont {J.~D.}\ \bibnamefont
  {Barrow}}\ and\ \bibinfo {author} {\bibfnamefont {A.~R.}\ \bibnamefont
  {Liddle}},\ }\href {\doibase 10.1103/PhysRevD.47.R5219} {\bibfield  {journal}
  {\bibinfo  {journal} {Phys.Rev.}\ }\textbf {\bibinfo {volume} {D47}},\
  \bibinfo {pages} {5219} (\bibinfo {year} {1993})},\ \Eprint
  {http://arxiv.org/abs/astro-ph/9303011} {arXiv:astro-ph/9303011 [astro-ph]}
  \BibitemShut {NoStop}%
%%CITATION = ASTRO-PH/9303011;%%
\bibitem [{\citenamefont {Barrow}\ \emph {et~al.}(2006)\citenamefont {Barrow},
  \citenamefont {Liddle},\ and\ \citenamefont {Pahud}}]{intinfl3}%
  \BibitemOpen
  \bibfield  {author} {\bibinfo {author} {\bibfnamefont {J.~D.}\ \bibnamefont
  {Barrow}}, \bibinfo {author} {\bibfnamefont {A.~R.}\ \bibnamefont {Liddle}},
  \ and\ \bibinfo {author} {\bibfnamefont {C.}~\bibnamefont {Pahud}},\ }\href
  {\doibase 10.1103/PhysRevD.74.127305} {\bibfield  {journal} {\bibinfo
  {journal} {Phys.Rev.}\ }\textbf {\bibinfo {volume} {D74}},\ \bibinfo {pages}
  {127305} (\bibinfo {year} {2006})},\ \Eprint
  {http://arxiv.org/abs/astro-ph/0610807} {arXiv:astro-ph/0610807 [astro-ph]}
  \BibitemShut {NoStop}%
%%CITATION = ASTRO-PH/0610807;%%
\bibitem [{\citenamefont {Barrow}(1995)}]{intinfl4}%
  \BibitemOpen
  \bibfield  {author} {\bibinfo {author} {\bibfnamefont {J.~D.}\ \bibnamefont
  {Barrow}},\ }\href {\doibase 10.1103/PhysRevD.51.2729} {\bibfield  {journal}
  {\bibinfo  {journal} {Phys.Rev.}\ }\textbf {\bibinfo {volume} {D51}},\
  \bibinfo {pages} {2729} (\bibinfo {year} {1995})}\BibitemShut {NoStop}%
%%CITATION = PHRVA,D51,2729;%%
\bibitem [{\citenamefont {{Barrow}}\ and\ \citenamefont
  {{Magueijo}}(2013)}]{inter-rainb}%
  \BibitemOpen
  \bibfield  {author} {\bibinfo {author} {\bibfnamefont {J.~D.}\ \bibnamefont
  {{Barrow}}}\ and\ \bibinfo {author} {\bibfnamefont {J.}~\bibnamefont
  {{Magueijo}}},\ }\href {\doibase 10.1103/PhysRevD.88.103525} {\bibfield
  {journal} {\bibinfo  {journal} {\prd}\ }\textbf {\bibinfo {volume} {88}},\
  \bibinfo {eid} {103525} (\bibinfo {year} {2013})},\ \Eprint
  {http://arxiv.org/abs/1310.2072} {arXiv:1310.2072 [astro-ph.CO]} \BibitemShut
  {NoStop}%
\bibitem [{\citenamefont {Ade}\ \emph {et~al.}(2013)\citenamefont {Ade} \emph
  {et~al.}}]{Planck}%
  \BibitemOpen
  \bibfield  {author} {\bibinfo {author} {\bibfnamefont {P.}~\bibnamefont
  {Ade}} \emph {et~al.} (\bibinfo {collaboration} {Planck Collaboration}),\
  }\href@noop {} {\  (\bibinfo {year} {2013})},\ \Eprint
  {http://arxiv.org/abs/1303.5076} {arXiv:1303.5076 [astro-ph.CO]} \BibitemShut
  {NoStop}%
%%CITATION = ARXIV:1303.5076;%%
\bibitem [{\citenamefont {Amelino-Camelia}\ \emph
  {et~al.}(2013{\natexlab{d}})\citenamefont {Amelino-Camelia}, \citenamefont
  {Arzano}, \citenamefont {Gubitosi},\ and\ \citenamefont
  {Magueijo}}]{measure}%
  \BibitemOpen
  \bibfield  {author} {\bibinfo {author} {\bibfnamefont {G.}~\bibnamefont
  {Amelino-Camelia}}, \bibinfo {author} {\bibfnamefont {M.}~\bibnamefont
  {Arzano}}, \bibinfo {author} {\bibfnamefont {G.}~\bibnamefont {Gubitosi}}, \
  and\ \bibinfo {author} {\bibfnamefont {J.}~\bibnamefont {Magueijo}},\
  }\href@noop {} {\  (\bibinfo {year} {2013}{\natexlab{d}})},\ \Eprint
  {http://arxiv.org/abs/1309.3999} {arXiv:1309.3999 [gr-qc]} \BibitemShut
  {NoStop}%
%%CITATION = ARXIV:1309.3999;%%
\bibitem [{\citenamefont {Calcagni}\ \emph {et~al.}(2013)\citenamefont
  {Calcagni}, \citenamefont {Eichhorn},\ and\ \citenamefont
  {Saueressig}}]{astrid}%
  \BibitemOpen
  \bibfield  {author} {\bibinfo {author} {\bibfnamefont {G.}~\bibnamefont
  {Calcagni}}, \bibinfo {author} {\bibfnamefont {A.}~\bibnamefont {Eichhorn}},
  \ and\ \bibinfo {author} {\bibfnamefont {F.}~\bibnamefont {Saueressig}},\
  }\href {\doibase 10.1103/PhysRevD.87.124028} {\bibfield  {journal} {\bibinfo
  {journal} {Phys.Rev.}\ }\textbf {\bibinfo {volume} {D87}},\ \bibinfo {pages}
  {124028} (\bibinfo {year} {2013})},\ \Eprint {http://arxiv.org/abs/1304.7247}
  {arXiv:1304.7247 [hep-th]} \BibitemShut {NoStop}%
%%CITATION = ARXIV:1304.7247;%%
\bibitem [{\citenamefont {Ambjorn}\ \emph
  {et~al.}(2005{\natexlab{b}})\citenamefont {Ambjorn}, \citenamefont
  {Jurkiewicz},\ and\ \citenamefont {Loll}}]{Ambjorn:2005db}%
  \BibitemOpen
  \bibfield  {author} {\bibinfo {author} {\bibfnamefont {J.}~\bibnamefont
  {Ambjorn}}, \bibinfo {author} {\bibfnamefont {J.}~\bibnamefont {Jurkiewicz}},
  \ and\ \bibinfo {author} {\bibfnamefont {R.}~\bibnamefont {Loll}},\ }\href
  {\doibase 10.1103/PhysRevLett.95.171301} {\bibfield  {journal} {\bibinfo
  {journal} {Phys.Rev.Lett.}\ }\textbf {\bibinfo {volume} {95}},\ \bibinfo
  {pages} {171301} (\bibinfo {year} {2005}{\natexlab{b}})},\ \Eprint
  {http://arxiv.org/abs/hep-th/0505113} {arXiv:hep-th/0505113 [hep-th]}
  \BibitemShut {NoStop}%
%%CITATION = HEP-TH/0505113;%%
\bibitem [{\citenamefont {Ambjorn}\ \emph {et~al.}(1998)\citenamefont
  {Ambjorn}, \citenamefont {Boulatov}, \citenamefont {Nielsen}, \citenamefont
  {Rolf},\ and\ \citenamefont {Watabiki}}]{Ambjorn:1997jf}%
  \BibitemOpen
  \bibfield  {author} {\bibinfo {author} {\bibfnamefont {J.}~\bibnamefont
  {Ambjorn}}, \bibinfo {author} {\bibfnamefont {D.}~\bibnamefont {Boulatov}},
  \bibinfo {author} {\bibfnamefont {J.~L.}\ \bibnamefont {Nielsen}}, \bibinfo
  {author} {\bibfnamefont {J.}~\bibnamefont {Rolf}}, \ and\ \bibinfo {author}
  {\bibfnamefont {Y.}~\bibnamefont {Watabiki}},\ }\href {\doibase
  10.1088/1126-6708/1998/02/010} {\bibfield  {journal} {\bibinfo  {journal}
  {JHEP}\ }\textbf {\bibinfo {volume} {9802}},\ \bibinfo {pages} {010}
  (\bibinfo {year} {1998})},\ \Eprint {http://arxiv.org/abs/hep-th/9801099}
  {arXiv:hep-th/9801099 [hep-th]} \BibitemShut {NoStop}%
%%CITATION = HEP-TH/9801099;%%
\bibitem [{\citenamefont {Horava}(2009{\natexlab{c}})}]{Horava:2009if}%
  \BibitemOpen
  \bibfield  {author} {\bibinfo {author} {\bibfnamefont {P.}~\bibnamefont
  {Horava}},\ }\href {\doibase 10.1103/PhysRevLett.102.161301} {\bibfield
  {journal} {\bibinfo  {journal} {Phys.Rev.Lett.}\ }\textbf {\bibinfo {volume}
  {102}},\ \bibinfo {pages} {161301} (\bibinfo {year} {2009}{\natexlab{c}})},\
  \Eprint {http://arxiv.org/abs/0902.3657} {arXiv:0902.3657 [hep-th]}
  \BibitemShut {NoStop}%
%%CITATION = ARXIV:0902.3657;%%
\bibitem [{\citenamefont {Lauscher}\ and\ \citenamefont
  {Reuter}(2005{\natexlab{b}})}]{Lauscher:2005qz}%
  \BibitemOpen
  \bibfield  {author} {\bibinfo {author} {\bibfnamefont {O.}~\bibnamefont
  {Lauscher}}\ and\ \bibinfo {author} {\bibfnamefont {M.}~\bibnamefont
  {Reuter}},\ }\href {\doibase 10.1088/1126-6708/2005/10/050} {\bibfield
  {journal} {\bibinfo  {journal} {JHEP}\ }\textbf {\bibinfo {volume} {0510}},\
  \bibinfo {pages} {050} (\bibinfo {year} {2005}{\natexlab{b}})},\ \Eprint
  {http://arxiv.org/abs/hep-th/0508202} {arXiv:hep-th/0508202 [hep-th]}
  \BibitemShut {NoStop}%
%%CITATION = HEP-TH/0508202;%%
\bibitem [{\citenamefont {Kommu}(2012)}]{Kommu:2011wd}%
  \BibitemOpen
  \bibfield  {author} {\bibinfo {author} {\bibfnamefont {R.}~\bibnamefont
  {Kommu}},\ }\href {\doibase 10.1088/0264-9381/29/10/105003} {\bibfield
  {journal} {\bibinfo  {journal} {Class.Quant.Grav.}\ }\textbf {\bibinfo
  {volume} {29}},\ \bibinfo {pages} {105003} (\bibinfo {year} {2012})},\
  \Eprint {http://arxiv.org/abs/1110.6875} {arXiv:1110.6875 [gr-qc]}
  \BibitemShut {NoStop}%
%%CITATION = ARXIV:1110.6875;%%
\bibitem [{\citenamefont {Giasemidis}\ \emph {et~al.}(2012)\citenamefont
  {Giasemidis}, \citenamefont {Wheater},\ and\ \citenamefont
  {Zohren}}]{Giasemidis:2012rf}%
  \BibitemOpen
  \bibfield  {author} {\bibinfo {author} {\bibfnamefont {G.}~\bibnamefont
  {Giasemidis}}, \bibinfo {author} {\bibfnamefont {J.~F.}\ \bibnamefont
  {Wheater}}, \ and\ \bibinfo {author} {\bibfnamefont {S.}~\bibnamefont
  {Zohren}},\ }\href {\doibase 10.1088/1751-8113/45/35/355001} {\bibfield
  {journal} {\bibinfo  {journal} {J.Phys.}\ }\textbf {\bibinfo {volume}
  {A45}},\ \bibinfo {pages} {355001} (\bibinfo {year} {2012})},\ \Eprint
  {http://arxiv.org/abs/1202.6322} {arXiv:1202.6322 [hep-th]} \BibitemShut
  {NoStop}%
%%CITATION = ARXIV:1202.6322;%%
\bibitem [{\citenamefont {Modesto}(2009{\natexlab{b}})}]{Modesto:2008jz}%
  \BibitemOpen
  \bibfield  {author} {\bibinfo {author} {\bibfnamefont {L.}~\bibnamefont
  {Modesto}},\ }\href {\doibase 10.1088/0264-9381/26/24/242002} {\bibfield
  {journal} {\bibinfo  {journal} {Class.Quant.Grav.}\ }\textbf {\bibinfo
  {volume} {26}},\ \bibinfo {pages} {242002} (\bibinfo {year}
  {2009}{\natexlab{b}})},\ \Eprint {http://arxiv.org/abs/0812.2214}
  {arXiv:0812.2214 [gr-qc]} \BibitemShut {NoStop}%
%%CITATION = ARXIV:0812.2214;%%
\bibitem [{\citenamefont {Calcagni}(2011{\natexlab{b}})}]{Calcagni:2010pa}%
  \BibitemOpen
  \bibfield  {author} {\bibinfo {author} {\bibfnamefont {G.}~\bibnamefont
  {Calcagni}},\ }\href {\doibase 10.1016/j.physletb.2011.01.063} {\bibfield
  {journal} {\bibinfo  {journal} {Phys.Lett.}\ }\textbf {\bibinfo {volume}
  {B697}},\ \bibinfo {pages} {251} (\bibinfo {year} {2011}{\natexlab{b}})},\
  \Eprint {http://arxiv.org/abs/1012.1244} {arXiv:1012.1244 [hep-th]}
  \BibitemShut {NoStop}%
%%CITATION = ARXIV:1012.1244;%%
\bibitem [{\citenamefont {Dodelson}(2003)}]{Dodelson:2003}%
  \BibitemOpen
  \bibfield  {author} {\bibinfo {author} {\bibfnamefont {S.}~\bibnamefont
  {Dodelson}},\ }\href
  {http://www.bibsonomy.org/bibtex/22a950650d49f9c18aced086d2231d499/richterek}
  {\emph {\bibinfo {title} {{Modern Cosmology}}}}\ (\bibinfo  {publisher}
  {Academic Press, Elsevier Science},\ \bibinfo {year} {2003})\BibitemShut
  {NoStop}%
\bibitem [{\citenamefont {Lu}\ and\ \citenamefont {Piao}(2010)}]{Lu:2009he}%
  \BibitemOpen
  \bibfield  {author} {\bibinfo {author} {\bibfnamefont {Y.}~\bibnamefont
  {Lu}}\ and\ \bibinfo {author} {\bibfnamefont {Y.-S.}\ \bibnamefont {Piao}},\
  }\href {\doibase 10.1142/S0218271810018074} {\bibfield  {journal} {\bibinfo
  {journal} {Int.J.Mod.Phys.}\ }\textbf {\bibinfo {volume} {D19}},\ \bibinfo
  {pages} {1905} (\bibinfo {year} {2010})},\ \Eprint
  {http://arxiv.org/abs/0907.3982} {arXiv:0907.3982 [hep-th]} \BibitemShut
  {NoStop}%
%%CITATION = ARXIV:0907.3982;%%
\bibitem [{\citenamefont {{Mukhanov}}\ \emph {et~al.}(1992)\citenamefont
  {{Mukhanov}}, \citenamefont {{Feldman}},\ and\ \citenamefont
  {{Brandenberger}}}]{Mukhanov:1990me}%
  \BibitemOpen
  \bibfield  {author} {\bibinfo {author} {\bibfnamefont {V.~F.}\ \bibnamefont
  {{Mukhanov}}}, \bibinfo {author} {\bibfnamefont {H.~A.}\ \bibnamefont
  {{Feldman}}}, \ and\ \bibinfo {author} {\bibfnamefont {R.~H.}\ \bibnamefont
  {{Brandenberger}}},\ }\href {\doibase 10.1016/0370-1573(92)90044-Z}
  {\bibfield  {journal} {\bibinfo  {journal} {Phys.Rep.}\ }\textbf {\bibinfo
  {volume} {215}},\ \bibinfo {pages} {203} (\bibinfo {year}
  {1992})}\BibitemShut {NoStop}%
\bibitem [{\citenamefont {Chen}(2010)}]{Chen:2010xka}%
  \BibitemOpen
  \bibfield  {author} {\bibinfo {author} {\bibfnamefont {X.}~\bibnamefont
  {Chen}},\ }\href {\doibase 10.1155/2010/638979} {\bibfield  {journal}
  {\bibinfo  {journal} {Adv.Astron.}\ }\textbf {\bibinfo {volume} {2010}},\
  \bibinfo {pages} {638979} (\bibinfo {year} {2010})},\ \Eprint
  {http://arxiv.org/abs/1002.1416} {arXiv:1002.1416 [astro-ph.CO]} \BibitemShut
  {NoStop}%
%%CITATION = ARXIV:1002.1416;%%
\end{thebibliography}%

\end{document}